\documentclass{jfm}

\usepackage{graphicx}
\usepackage{natbib}
\usepackage{math_pack}
\usepackage{color}
\usepackage{subfigure}
\usepackage{amsmath}
\usepackage{amssymb}
\usepackage{math_pack}

\ifCUPmtlplainloaded \else
  \checkfont{eurm10}
  \iffontfound
    \IfFileExists{upmath.sty}
      {\typeout{^^JFound AMS Euler Roman fonts on the system,
                   using the 'upmath' package.^^J}%
       \usepackage{upmath}}
      {\typeout{^^JFound AMS Euler Roman fonts on the system, but you
                   dont seem to have the}%
       \typeout{'upmath' package installed. JFM.cls can take advantage
                 of these fonts,^^Jif you use 'upmath' package.^^J}%
      }
  \else
  \fi
\fi

\ifCUPmtlplainloaded \else
  \checkfont{msam10}
  \iffontfound
    \IfFileExists{amssymb.sty}
      {\typeout{^^JFound AMS Symbol fonts on the system, using the
                'amssymb' package.^^J}%
       \usepackage{amssymb}%
         \let\leq=\leqslant
         \let\geq=\geqslant
      }{}
  \fi
\fi

\ifCUPmtlplainloaded \else
  \IfFileExists{amsbsy.sty}
    {\typeout{^^JFound the 'amsbsy' package on the system, using it.^^J}%
     \usepackage{amsbsy}}
    {\providecommand\boldsymbol[1]{\mbox{\boldmath $##1$}}}
\fi

\newcommand\omegab{\boldsymbol{\omega}}
\newcommand\taub{\boldsymbol{\tau}}

\newcommand\AR{\mbox{\textit{AR}}}            

\newsavebox{\astrutbox}
\sbox{\astrutbox}{\rule[-5pt]{0pt}{20pt}}

\definecolor{cinnamon}{rgb}{0.82, 0.41, 0.12}

\title[Clustering and increased settling speed of oblate particles]
{Clustering and increased settling speed of oblate particles at finite Reynolds number}

\author[W. Fornari, M. N. Ardekani and L. Brandt]%
{Walter Fornari$^1$%
  \thanks{Email address for correspondence: fornari@mech.kth.se},
Mehdi Niazi Ardekani$^1$ \\
and Luca Brandt$^1$}

\affiliation{$^1$Linn\'e Flow Centre and Swedish e-Science Research Centre (SeRC), \\ KTH Mechanics,
SE-100 44 Stockholm, Sweden\\[\affilskip]}

\pubyear{2010}
\volume{650}
\pagerange{119--126}
\date{?; revised ?; accepted ?. - To be entered by editorial office}
\begin{document}

\maketitle

\begin{abstract}

We study the settling of rigid oblates in quiescent fluid using interface-resolved Direct Numerical Simulations. In 
particular, an immersed boundary method is used to account for the dispersed solid phase together with lubrication correction and collision models to account for short-range particle-particle interactions.
 We consider semi-dilute suspensions of oblate particles with aspect ratio
$\AR=1/3$ and solid volume fractions $\phi=0.5\%-10\%$. The solid-to-fluid density ratio $R=1.5$ and the Galileo number (i.e. the ratio between buoyancy and 
viscous forces) based on the diameter of a sphere with equivalent volume $Ga=60$. 
With this choice of parameters, 
an isolated oblate falls vertically with a steady wake with its broad side perpendicular to the gravity direction. At this $Ga$, the mean settling speed of spheres is a 
decreasing function of the volume $\phi$ and is always smaller than the terminal velocity of the isolated particle, $V_t$. 
On the contrary, we show here that the mean settling speed of oblate particles increases with $\phi$ in dilute conditions and is $33\%$ 
larger than $V_t$. At higher concentrations, the mean settling speed decreases becoming smaller than the terminal velocity $V_t$ 
between $\phi=5\%$ and $10\%$. The increase of the mean settling speed is due to the formation of particle 
clusters that for $\phi=0.5\%-1\%$ appear as columnar-like structures. From the 
pair-distribution function we observe that it is most probable to find particle-pairs almost vertically 
aligned. However, the pair-distribution function is non-negligible all around the reference particle indicating 
that there is a substantial amount of clustering at radial distances between 2 and $6c$ (with $c$ the polar 
radius of the oblate). 
Above $\phi=5\%$, the hindrance becomes the dominant effect, and the mean settling speed decreases below $V_t$.
As the particle concentration increases, the mean particle orientation changes and the mean pitch angle (the 
angle between the particle axis of symmetry and gravity) increases from $23^o$ to $47^o$.
\end{abstract}


\section{Introduction}

There is a wide range of environmental processes and industrial applications that involve suspensions of particles 
settling under gravity. Among these we recall the pollutant transport in underground water, soot particle 
dispersion, fluidized beds and the settling of micro-organisms such as plankton, rain droplets and snow.\\
Often, these applications involve a large number of particles settling in quiescent fluids and despite the large 
number of studies on the topic, the understanding of this complex phenomenon is still far from clear. 
Sedimentation depends indeed on a wide range of parameters. Particles may differ in density, shape, size and 
stiffness, and real suspensions are hardly monodispersed. 
In the present work, we focus on the effects due to particle shape. In particular, we consider suspensions of 
buoyant oblate particles of fixed aspect ratio ($\AR=1/3$), and show how particle orientation leads to different 
dynamics and microstructures in comparison to the ideal case of spherical particles.

If we limit our attention to the case of an isolated 
rigid sphere, it is known that the settling speed depends on the solid-to-fluid density ratio, $R$, and the Galileo 
number $Ga$, namely the ratio between buoyancy and viscous forces acting on the particle. Even in this 
two-parameter space a variety of particle path regimes are encountered, involving vertical, oblique, 
time-periodic oscillating, zig-zagging, helical and chaotic motion as shown numerically and experimentally by 
\citet{jenny2004,horowitz2010}. The a-priori estimation of the particle terminal falling velocity, $V_t$ is also 
non-trivial. With the assumption of Stokes flow in an unbounded quiescent fluid, it is found that the Stokes 
terminal velocity $V_s$ is function of the sphere radius, $a$, the density ratio, $R$, the magnitude of the 
gravitational acceleration $g$ and the kinematic viscosity of the fluid $\nu$ \citep{guazzelli2011}. However, 
when the Reynolds number of the settling particle ($Re_t=V_t d/\nu$) becomes finite, 
the fore-aft symmetry of the fluid flow around the particle is broken leading to 
the generation of a rear wake. As previously mentioned, the terminal speed depends on $R$ and more pronouncedly  on 
$Ga$, but no theoretical formula that relates these quantities exists. Up to date, only formulae that make use 
of empirical relations for the drag coefficient of an isolated sphere, $C_D$, are available \citep{schil1935,
clift2005,yin2007}. However, these formulae relate only the terminal Reynolds number $Re_t$ to the Galileo 
number $Ga$, neglecting the dependence on the density ratio $R$.

When particle suspensions are considered, the scenario is further complicated by hydrodynamic and particle-particle 
interactions. A most relevant effect occur when a sphere is entrained in the wake of another particle of comparable size 
settling at finite $Re_t$, as the particle behind will accelerate towards the leading particle. The particles will hence touch and 
finally the rear particle will tumble laterally. This phenomenon is denoted as  \emph{drafting-kissing-tumbling} 
of a particle-pair \citep{fortes1987}; during the draft phase the rear particle reaches speeds larger 
than the terminal velocity $V_t$. The extent of the increase of the rear particle speed with respect to $V_t$ 
depends on $Ga$. 

Generally speaking, the mean settling speed of a suspension of particles, $\langle V_z \rangle$, is also a 
function of the solid volume fraction $\phi$. For very dilute suspensions under the assumption of Stokes flow, 
\citet{hasimoto1959} and later \citet{sangani1982} obtained expressions for the drag force exerted by the fluid 
on three different cubic arrays of settling spheres. A different expression was instead found by 
\citet{batchelor1972} who used a different approach based on conditional probability arguments. All these 
formulae relate the mean settling speed directly to the solid volume fraction $\phi$ but are unable to properly 
predict $\langle V_z \rangle$ for semi-dilute and dense suspensions. For such suspensions, the empirical 
formula proposed by \citet{richardson1954} is probably the most used. This was obtained from experimental results in creeping 
flow conditions and relates the mean settling speed normalized by the Stokes terminal velocity to the solid 
volume fraction $\phi$, via a power-law. More specifically, the mean settling speed of the suspension $\langle 
V_z \rangle$ is a decreasing function of $\phi$ and is always smaller than $V_s$. This formula has been shown to 
be accurate also for concentrated suspensions and for low Reynolds numbers $Re_t$. A wide number of more recent 
studies have been devoted to the improvement of this empirical formula to account for larger $Re_t$. Among these we 
recall the experimental studies by \citet{garside1977,di1999} and the numerical study by \citet{yin2007}. These 
authors showed that the power-law exponent is a non-linear function of the Reynolds number $Re_t$, and that a 
correction coefficient should be introduced.

The mean settling speed $\langle V_z \rangle$ decreases with $\phi$ due to the hindrance effect 
\citep{climent2003,guazzelli2011}. In a batch sedimentation system, the fixed bottom of the container forces the 
fluid to move in the opposite direction such that the flux of the particle-fluid mixture remains zero. Hindrance 
becomes more pronounced as the volume fraction $\phi$ increases, leading a monotonic decrease of $\langle V_z 
\rangle$ with respect to $V_t$. At large $Re_t$, however, the suspension behavior is further complicated by the 
particle-particle hydrodynamic interactions. In our previous work, we have studied semi-dilute suspensions 
($\phi=0.5\%-1\%$) of spheres with density ratio $R=1.02$ and $Ga=145$ \citep{fornari2015} and have found that 
drafting-kissing-tumbling events are indeed frequent, with the involved particles reaching speeds more than 
twice the mean $\langle V_z \rangle$. It was estimated that without these intermittent events the mean settling speed, 
$\langle V_z \rangle$, would be smaller by about $3\%$.

For suspensions of spheres with $\phi=0.5\%$, $R=1.5$ and a larger value $Ga=178$, \citet{uhlmann2014} found that 
particle clusters form. These clusters settle faster than $V_t$ and as a result, the mean settling speed $\langle V_z 
\rangle$ increases by $12\%$ with respect to the terminal speed of an isolated particle, $V_t$. The formation of clusters is 
related to the steady oblique motion observed for isolated spheres with $R=1.5$ and $Ga=178$. Indeed, at a lower 
$Ga=121$, for wihch an isolated sphere exhibits a steady vertical motion, no clustering is observed. An increased mean 
settling speed at large $Ga$ was also observed by \citet{zaidi2014,fornari2016b}. Recently, these results were 
also confirmed experimentally by \citet{huisman2016} who also observed the formation of a columnar structure of 
spheres at high $Ga$.

In the past few years, numerical investigations were also devoted to the study of the sedimentation of 
suspensions of finite-size spheres in stratified environments \citep{doost2015}, in homogeneous isotropic 
\citep{chouippe2015,fornari2015,fornari2016b} and shear turbulence \citep{tanaka2015}. The case of finite-size 
spherical bubbles rising in vertical turbulent channel flow has been considered by \citet{santarelli2015,
santarelli2016}.

When considering non-spherical particles, the sedimentation process is further complicated as the particle 
orientation plays a role in the dynamics. \citet{feng1994} performed two-dimensional numerical simulations of 
settling elliptic particles to show that in stable conditions an elliptic particle always falls with its 
long axis perpendicular to gravity. Three-dimensional oblates settling in steady motion at low $Re_t$ also display
the symmetry axis in the gravity direction. However, increasing $R$ or $Ga$ the system becomes 
unstable and disc-like particles are observed to oscillate horizontally. As explained by \citet{magnaudet2007,
ern2012}, the path instability of spheroidal particles is closely related to their wake instability. Indeed, the 
release of vortices in the wake of a spheroidal particle is modified as soon as the angle between the particle 
symmetry axis and the velocity direction is changed. Therefore, the ensuing wake instability is also strongly 
related to the particle aspect ratio $\AR$. A complete parametric study on disc-shaped cylinders and oblates with 
different aspect and density ratios falling under gravity was performed by \citet{chrust2012}.

Recently we extended the immersed boundary method (IBM) of \citet{breugem2012} to account for 
ellipsoidal particles \citep{ardekani2016}. We have shown that above a threshold $Ga$, oblate particles perform a 
zigzagging motion whereas prolate particles rotate around the vertical axis with their broad side facing the 
falling direction. The threshold $Ga$ is shown to decrease as the aspect ratio departs from $1$. Particle-pair 
interactions were also studied. It has been found that the drafting-kissing-tumbling is modified with respect to 
the case of settling spheres. In particular, for two oblate particles with $\AR=1/3$ and $Ga=80$, 
the tumbling part is suppressed and the particles fall together with a mean speed that is substantially 
larger than $V_t$. Also,
 spheroidal particles are attracted in the wake of a leading particle from larger lateral 
distances than in the case of spheres. The absence of the tumbling phase was also found experimentally for pairs of falling disks with $\AR=1/6$ by \citet{brosse2011}.

Much less is known about the sedimentation of suspensions of spheroidal particles. A pioneering study is that by 
\citet{fonseca}, who studied numerically the settling of suspensions of oblate ellipsoids with $\AR \simeq 0.27$ 
and $R=4$ at the relatively low Reynolds numbers, $Re_t= 0.04$ and $7$. Volume fractions up to $\phi \sim 0.2$ were considered. At the smallest 
$Re_t$, these authors found a local maxima of the mean settling speed $\langle V_z \rangle$ at $\phi=0.05$, that 
is however smaller than the settling speed of an isolated oblate, $V_t$. On the other hand, at $Re_t=7$ it was 
found that $\langle V_z \rangle \simeq 1.1 V_t$ for $\phi \sim 1\%$. 

In the present study, we investigate the sedimentation of semi-dilute suspensions of oblate particles 
at finite Reynolds number $Re_t$. In particular, we consider particles with aspect ratio $\AR=1/3$, density ratio 
$R=1.5$ and Galileo number (based on the diameter of a sphere with an equivalent volume) $Ga=60$. With this choice of 
parameters, a single oblate falls steadily with its broad side perpendicular to gravity and with a terminal 
Reynolds number of approximately $40$. Four solid volume fractions of $\phi=0.5\%, 1\%, 5\%$ and $10\%$ are 
studied. We find that differently from spheres of equal $Ga$, the mean settling speed of the suspension, $\langle 
V_z \rangle$, first increases with $\phi$, and is therefore larger than the terminal velocity of a single particle, $V_t$. The mean 
settling speed decreases for $\phi > 0.5\%$ and becomes smaller than $V_t$ between $\phi=5\%$ and $10\%$. In 
this range of $\phi$, a power-law fit similar to that by \citet{richardson1954} is proposed. We then show that 
the enhancement of $\langle V_z \rangle/V_t$ at low $\phi$ is related to the formation of a columnar structure of 
particles. Within this structure, intense particle clustering is observed. For $\phi=0.5\%-1\%$, the particle
pair-distribution function is found to be high in the range $r \in [2c, 6c]$ and between $\psi \simeq 2^o-80^o$, with maximum 
values at $r=2.02c$ and $\psi=17^o-10^o$ (with $c$ the polar radius of the oblate and $\psi$ the polar angle with 
respect to the direction of gravity). Hence, particles are almost vertically piled-up at low $\phi$ as also shown by the order 
parameter. At higher $\phi$ the amount of clustering is reduced. 
We also show that the mean particle orientation (computed as the cosine of the angle between the particle symmetry 
axis and gravity) decreases with $\phi$. A power-law fit in terms of $\phi$ is also 
proposed. The particle mean pitch angle with respect to the horizontal plane increases 
with $\phi$, from $22.8^o$ ($\phi=0.5\%$) to $47^o$ ($\phi=10\%$). It should be noted that for an isolated oblate with $Ga = 60$ and $R=1.5$ the pitch 
angle is $0^o$. Finally, we calculate joint probability functions of settling speeds $V_z$ and orientation 
$|O_z|$. By means of conditioned averages we show that particles settling with larger speeds than the mean, 
$\langle V_z \rangle/V_t$, settle on average with higher pitch angles.

\section{Set-up and Methodology}\label{sec:method}

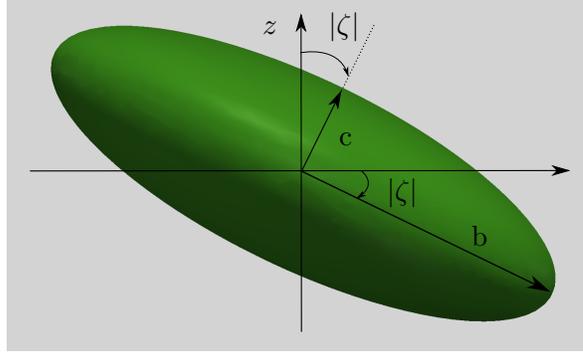
\begin{figure}
  \centering
  \subfigure{%
    \includegraphics[scale=0.15]{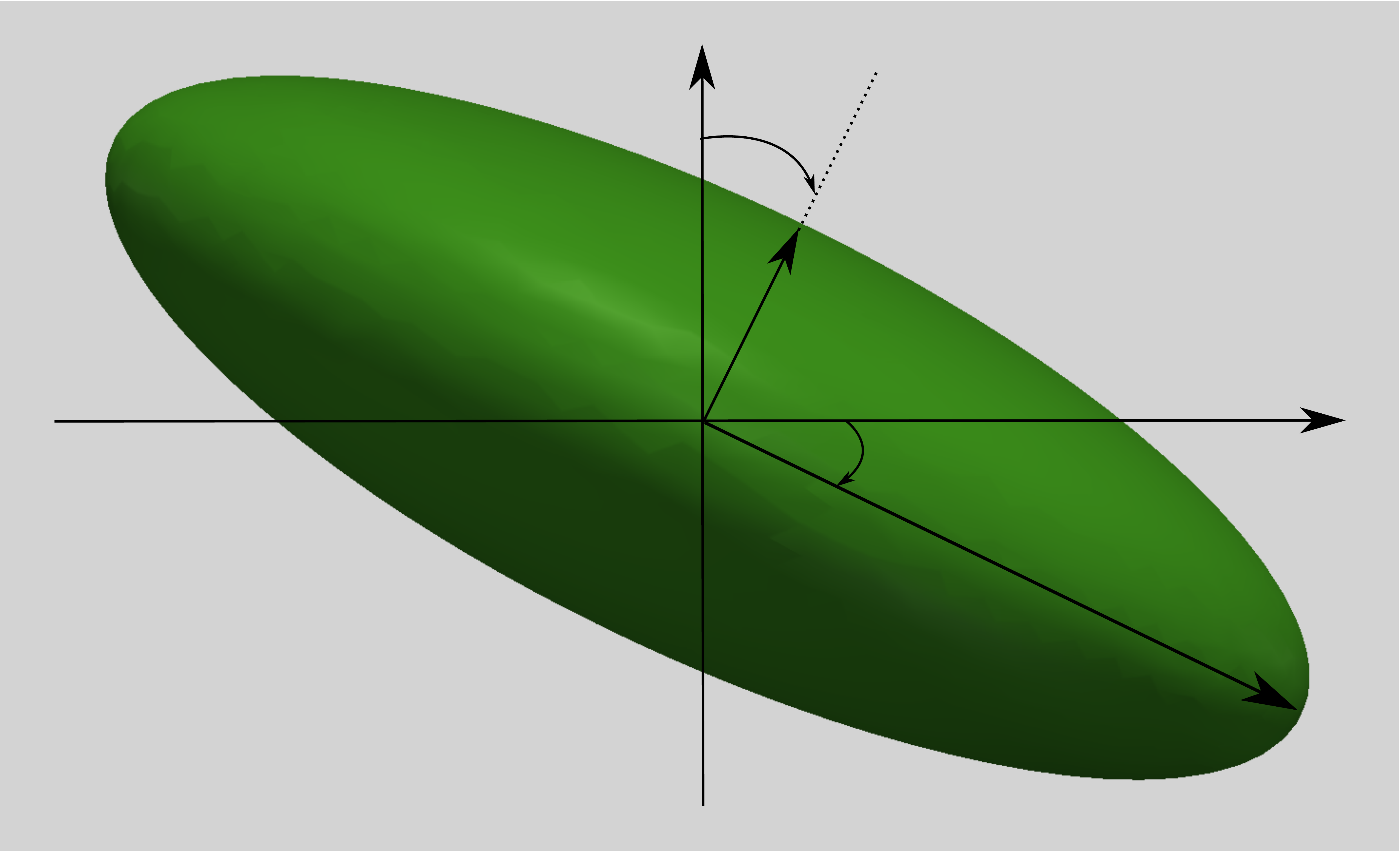}   
     \put(-96,78){{\large c}}
     \put(-46,40){{\large b}}
     \put(-78,57){{\large $|\zeta|$}}
     \put(-100,120){{\large $|\zeta|$}}
     \put(-125,120){{\large $z$}}
     }%
\caption{Definition of the equatorial and polar radii, $b$ and $c$, and of the pitch angle $|\zeta|$.}
\label{fig:plot}
\end{figure}

The sedimentation of semi-dilute suspensions of oblate particles is considered in a computational domain with 
periodic boundary conditions in the $x$, $y$ and $z$ directions for both the fluid and the particles, with gravity acting in the positive $z$ 
direction. Oblates with aspect ratio $\AR=1/3$ are considered. We name $b$ and $c$ the equatorial and polar radii 
of the ellipsoid. The computational box has size $20d\times20d\times160d$, being $d$ the diameter of a 
sphere with the same volume as the ellipsoidal particle. Four solid volume fractions are investigated, 
$\phi=0.5\%, 1\%, 5\%$ and $10\%$. These correspond to $611$, $1222$, $6111$ and $12222$ particles. Oblates are 
initially randomly distributed in the computational domain with zero angular and translational velocity, and with their 
orientation vector $[O_x,Oy,Oz]=[0, 0, 1]$. Hence, their broad side is perpendicular to gravity and the pitch 
angle (defined between their symmetry axis and gravity) is $0^o$. Note that this angle is equal to the angle between the 
plane defined by the equatorial radius $b$ and the (horizontal) $xy$ plane, as shown in figure~\ref{fig:plot}. Therefore 
we name it as pitch angle (i.e.\ when the pitch angle is larger than zero, the spheroid is inclined with respect to the 
horizontal plane). For comparisons, the case of an isolated oblate is also simulated as reference. 

We consider non-Brownian rigid oblate particles slightly heavier than the suspending fluid with density ratio 
$R=1.5$ and Galileo number (based on the diameter $d$ of the equivalent sphere)
\begin{equation}
Ga=\frac{\sqrt{\left(R-1\right) g d^3}}{\nu}=60
\end{equation}
As previously mentioned, this non-dimensional number quantifies the importance of the gravitational forces acting 
on the particle with respect to viscous forces. At this $Ga$ isolated spheres and oblates settle vertically with 
steady wakes. The Reynolds number based on the terminal falling speed of a single oblate is found to be 
$Re_t=38.7$. 

The simulations have been performed using the version of the immersed boundary method developed by 
\citet{breugem2012} and modified by \citet{ardekani2016} to account for ellipsoidal particles. With this approach, 
the coupling between the fluid and solid phases is fully modelled. The flow is evolved according to the 
incompressible Navier-Stokes equations, whereas the particle motion is governed by the Newton-Euler Lagrangian 
equations for the particle centroid linear and angular velocities
\begin{align}
\label{lin-vel}
\rho_p V_p \td{\vec u_p}{t} &= \oint_{\partial \mathcal{V}_p}^{} \vec \taub \cdot \vec n\, dS + \left(\rho_p - \rho_f\right) V_p \vec g\\
\label{ang-vel}
\td{I_p \vec \omegab_p}{t} &= \oint_{\partial \mathcal{V}_p}^{} \vec r \times \vec \taub \cdot \vec n\, dS
\end{align}
where $\rho_p$, $V_p$ and $I_p$ are the particle density, volume and moment of inertia; $\vec g$ is the 
gravitational acceleration; $\vec \taub = -p \vec I + 2\mu \vec E$ is the fluid stress, with $\vec E = 
\left(\grad \vec u_f + \grad \vec u_f^T \right)/2$ the deformation tensor; $\vec r$ is the distance vector from 
the center of the particle while $\bf{n}$ is the unity vector normal to the particle surface $\partial 
\mathcal{V}_p$. Dirichlet boundary conditions for the fluid phase are enforced on the particle surfaces as 
$\vec u_f|_{\partial \mathcal{V}_p} = \vec u_p + \vec \omegab_p \times \vec r$.\\
Using the immersed boundary method, the boundary condition at the moving fluid/solid interfaces is indirectly imposed by an 
additional force on the right-hand side of the Navier-Stokes equations. It is hence possible to discretize the 
computational domain with a fixed staggered mesh on which the fluid phase is evolved using a second-order 
finite-difference scheme together with a set of Lagrangian points, uniformly distributed on the surface of the particle to represent the interface. Time integration is  performed by a third-order Runge-Kutta scheme combined with 
pressure correction at each sub-step. When the distance between two particles becomes smaller than twice the mesh 
size, a lubrication model is used to correctly reproduce the interaction between the particles. In particular, the 
closest points on the surfaces of two ellipsoids are found. From these, the Gaussian radii of curvature are calculated, and these correspond to the radii of the best fitting spheres tangent to the given surface points. The 
lubrication model based on \citet{jeffrey1982} asymptotic solution for spheres of different size is then employed. 
Additionally, the soft-sphere model is used to account for normal and tangential collisions between the ellipsoids 
\citep{costa2015}. As for lubrication, collision forces are calculated for the best fitting spheres at the points 
of contact and are later transferred to the spheroids centres. 
More details and validations for the specific immersed boundary method used for ellipsoids can be found in 
\citet{ardekani2016}. Other validations specific to the immersed boundary method for spherical particles are found in \citet{breugem2012,
lambert2013,picano2015,fornari2015}.\\
A cubic mesh with approximately eight points per particle polar radius $c$ ($\sim 24$ points per equatorial 
radius $b$) is used for the results presented, which corresponds to $640 \times 640\times5120$ grid points in the computational domain and $3220$ Lagrangian points on the surface of each particle. 
Note finally that zero total volume flux is imposed in the simulations. 

For $\phi=0.5\%-1\%$, simulations were run for $182$ particle relaxation times defined using the equivalent 
diameter $d$ ($\tau_p = R d^2/(18 \nu)$). Defining as reference time the time it takes for an isolated oblate to 
fall over a distance equal to its polar radius, $c/V_t$, the simulation time corresponds to $2430 \, c/V_t$. 
For denser cases, the statistically steady-state condition is reached earlier and simulations were run for $72 \, 
\tau_p = 962 \, c/V_t$ ($\phi=5\%$) and $42 \, \tau_p = 561 \, c/V_t$ ($\phi=10\%$). Statistics are collected 
after $90 \, \tau_p = 1202 \, c/V_t$ for $\phi=0.5\%-1\%$, $24 \, \tau_p = 320 \, c/V_t$ for $\phi=5\%$, and $6 \, 
\tau_p = 80 \, c/V_t$ for $\phi=10\%$.

\section{Results}
\subsection{Settling speed and suspension microstructure}

The most striking result of our study is that semi-dilute suspensions of oblate particles with $Ga \gg 1$, settle 
on average substantially faster than isolated particles, about 33\% faster at $\phi=0.5\%$.  As the Galileo number $Ga$ increases, the effects of 
particle inertia and hydrodynamic interactions due to particle wakes become progressively more important 
overcoming the hindrance effect described above \citep{yin2007,guazzelli2011}. For spherical particles at moderate $Ga$, the hindrance leads to a 
power-law decay of the mean settling speed $\langle V_z \rangle$ with the volume fraction $\phi$. Hence, the mean settling speed $V_z$ is smaller than the terminal falling speed $V_t$ of an isolated sphere  for all volume fractions $\phi$. At 
low and moderate terminal Reynolds numbers, $Re_t=V_td/\nu \leq 20$, the hindrance effect is well-described by the 
modified Richardson \& Zaki empirical formula \citep{richardson1954}
\begin{equation}
\label{zaki}
\frac{\langle V_z \rangle}{V_t} = \kappa (1-\phi)^n
\end{equation}
where $n$ is an exponent that depends on $Re_t$ \citep{garside1977}
\begin{equation}
\label{gars}
\frac{5.1-n}{n-2.7}=0.1 Re_t^{0.9}
\end{equation}
while $\kappa$ is a correction coefficient for finite $Re_t$ that has been found to be in the range $0.8 
- 0.92$ \citep{di1999,yin2007}. Note that we have checked that we can get consistent values of $\kappa$ with our code. In particular, we 
have also performed simulations of suspensions of spheres with $Ga \sim 9$ settling under gravity, and found that $\kappa \sim 0.91$. 

\begin{figure}
  \centerline{\includegraphics[scale=0.5]{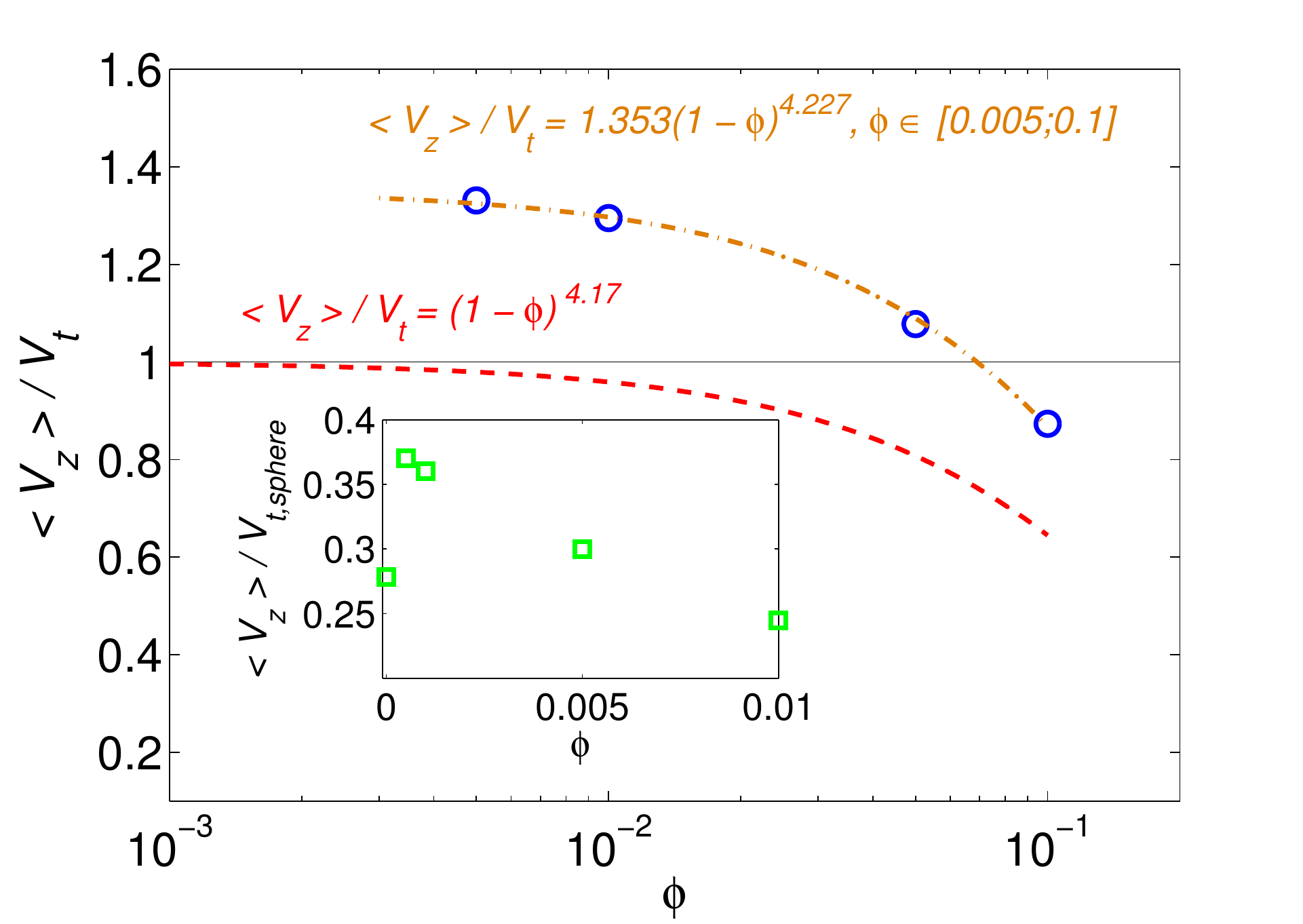}}
  \caption{Mean settling speed $\langle V_z \rangle/V_t$ as function of the solid volume fraction $\phi$. The 
dash-dotted line is a fit of $\langle V_z \rangle/V_t$ in the range $\phi \in [0.5; 10]\%$. The empirical fit 
proposed by Richardson and Zaki for settling spheres is also shown (dashed line). In the inset, the mean 
settling speed is normalized by the settling speed of an isolated spherical particle of same Galileo number 
$Ga$, based on the equivalent diameter.
\label{fig:mvel}}
\end{figure}

From the simulation of the isolated settling oblate we find that the terminal Reynolds number is 
$Re_t= V_t d_{eq}/\nu = 38.7$ (being $d_{eq}$ the equivalent diameter of a sphere with the same volume). At these 
$R$, $Ga$ and $Re_t$ the corresponding wake behind both spheres and oblates is steady and vertical 
\citep{bouchet2006,ardekani2016}. 
 For the case of oblique wakes past  isolated spheres, \citet{uhlmann2014}
have shown an increase of the mean settling speed
(these authors considered spheres with $R=1.5$ and $Ga=178$).
 The mean settling 
speed of the suspension ($\phi=0.5\%$) increases above $V_t$ by about $12\%$. This increase of $\langle V_z \rangle$ 
is due to the formation of particle clusters.

The results for the mean settling speed of the oblate suspension, $\langle V_z \rangle$, normalized by $V_t$, are 
shown in figure~\ref{fig:mvel}. The expected velocity predicted via the empirical fit~(\ref{zaki}) using 
$\kappa = 1$ and $n=4.17$ (obtained from equation \ref{gars} and $Re_t=38.7$) is also shown for comparison. 
In contrast to what expected for spheres, we find that the mean settling speed $\langle V_z \rangle$ is larger 
than $V_t$ for volume fractions approximatively  lower than $7\%$. For $\phi=0.5\%$ and $1\%$, $\langle V_z \rangle \sim 1.3 V_t$, and for 
$\phi=5\%$ $\langle V_z \rangle = 1.08 V_t$. Hence, as $\phi$ increases there is an initial increase of 
$\langle V_z \rangle/V_t$. However, the hindrance effect becomes progressively stronger 
above $\phi=0.5\%$, reducing $\langle V_z \rangle/V_t$; this observable becomes lower than $1$ for the largest volume fraction considered here. 
We fitted our data in the range $\phi \in [0.5\%-10\%]$
using equation~(\ref{zaki}) to find $\langle V_z \rangle/V_t = 1.353 (1-\phi)^{4.227}$. Clearly, this relation is valid 
only for this range of volume fractions and possibly for larger $\phi$. We leave as future work  
the study of more dilute cases to understand how $\langle V_z \rangle/V_t$ initially increases with $\phi$.

In the inset of figure~\ref{fig:mvel} we report the same data normalized by the terminal velocity of an isolated 
spherical particle with $Ga=60$. We see that oblates settle at a substantially slower rate than spheres for all 
$\phi$. For $\phi=0.5$, $1\%$, $\langle V_z \rangle \sim 0.36 V_{t,sphere}$.

\begin{figure}
  \centering
  \subfigure{%
    \includegraphics[scale=0.33]{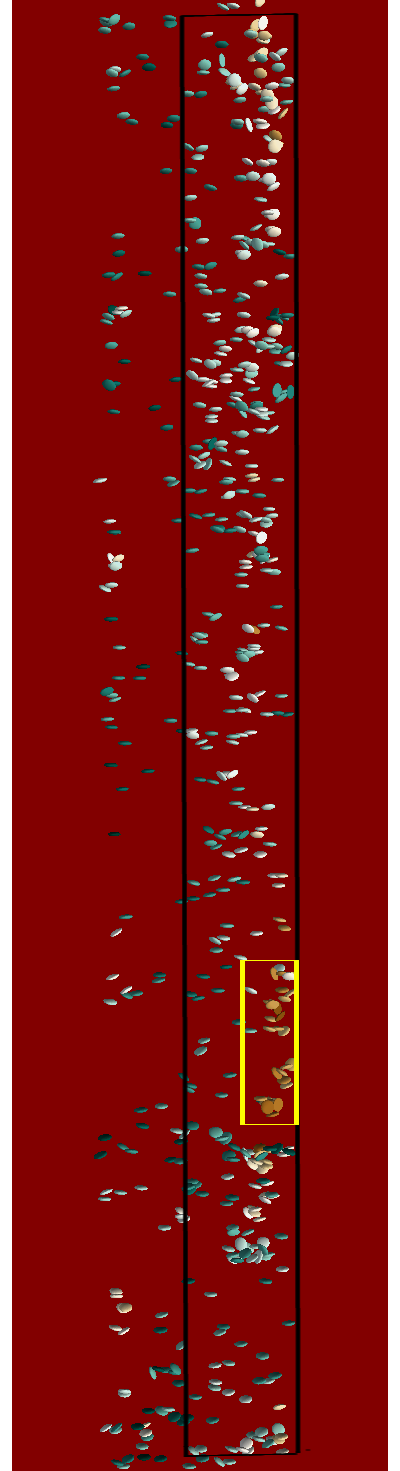}   
     \put(-124,470){{\large a)}}
     }%
  \subfigure{%
    \includegraphics[scale=0.33]{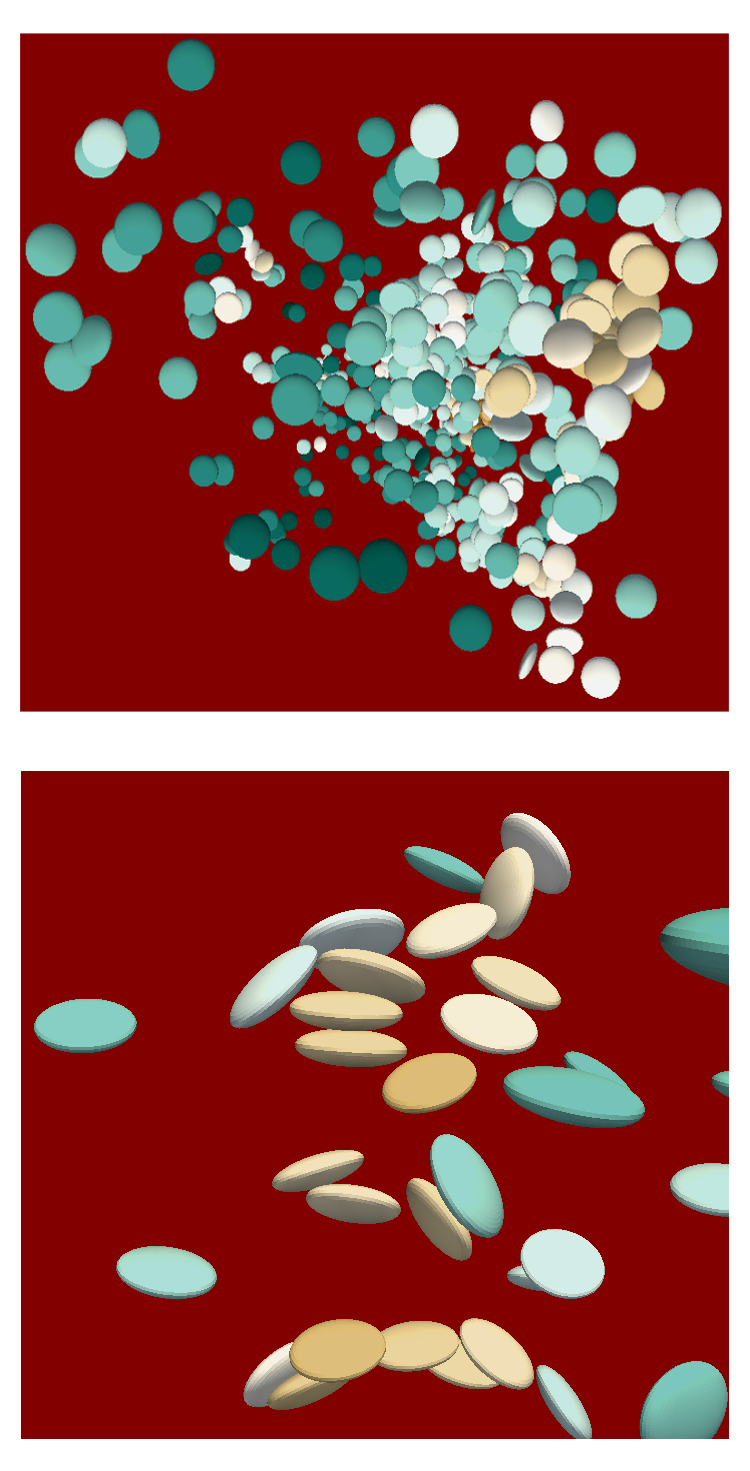}   
     \put(-130,480){{\large b)}}
     \put(-130,238){{\large c)}}
     }%
\caption{Instantaneous snapshots of oblate particles falling under gravity: a) front view; b) top view; c) zoom 
on a cluster of $3$ piled up particles.}
\label{fig:ivel}
\end{figure}

To understand the enhancement of $\langle V_z \rangle/V_t$ at moderate $\phi$, we first show in 
figure~\ref{fig:ivel}a) an instantaneous snapshot of the settling suspension at $\phi=0.5\%$. 
It can be seen that most particles are located on the right half of the snapshot (i.e. for $x \geq 0.5 L_x$) where they 
seem to form a columnar structure. As we will soon show, particles within this columnar structure fall on average 
faster than the more isolated particles and than the whole suspension. This peculiar particle distribution can also 
be observed in figure~\ref{fig:ivel}b), displaying a the top view of the same instantaneous configuration. These observations confirm the 
importance of particle-particle hydrodynamic interactions at finite inertia. Note that the smaller region delimited by a box  in 
figure~\ref{fig:ivel}a) highlights particles that fall approximately $3-3.5$ times faster than an isolated 
particle.

As mentioned above, \citet{ardekani2016} 
studied the settling of two isolated oblate particles at a similar $Ga$. While two spheres would undergo the 
so-called \emph{drafting-kissing-tumbling} phenomenon \citep{fortes1987}, \citet{ardekani2016} showed that the 
rear oblate accelerates in the wake of the front particle until it approaches it and almost perfectly sticks to 
it. The tumbling stage is therefore suppressed and the particles fall in contact. The particle-pair falls with 
a speed that is $1.5$ times the terminal velocity of an isolated oblate. In the same study, it was also found that the maximum 
radius of the collision (or entrainment area) for oblates of $\AR = 1/3$ is approximately $4$ times larger than 
that of spheres of equal $Ga$ for several different vertical separations. 
These results obtained for particle-pairs are reflected in  and determine the 
suspension behavior. From the close-up in figure~\ref{fig:ivel}c), we note that in a suspension with $\phi=0.5\%$ more than 
$3$ particles can pile up during drafting-kissing-tumbling events. These particle clusters generate strong wakes 
that, in turn, lead to the formation of the columnar-like structure. This columnar structure is observed also for 
larger $\phi$. Note that the $(x,y)$ location of the structure is purely random. 
 Similar columnar structures were observed also for spherical particles by \citet{uhlmann2014,huisman2016}, although for much larger $Ga=178$, while  these were not observed for spheres at $Ga=120$. 

\begin{figure}
  \centering
  \subfigure{%
    \includegraphics[scale=0.35]{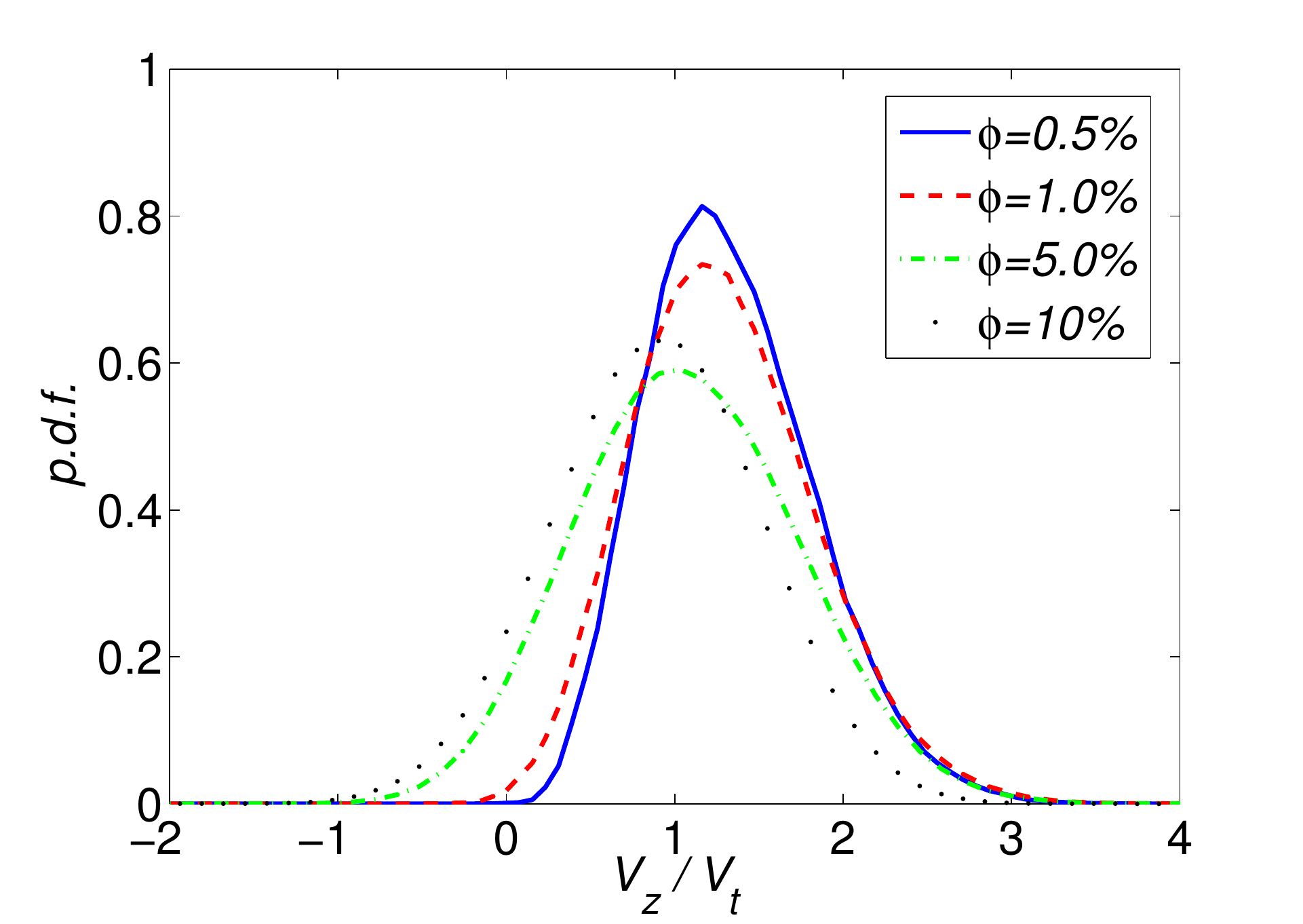}   
     \put(-194,120){{\large a)}}
     }%
  \subfigure{%
    \includegraphics[scale=0.35]{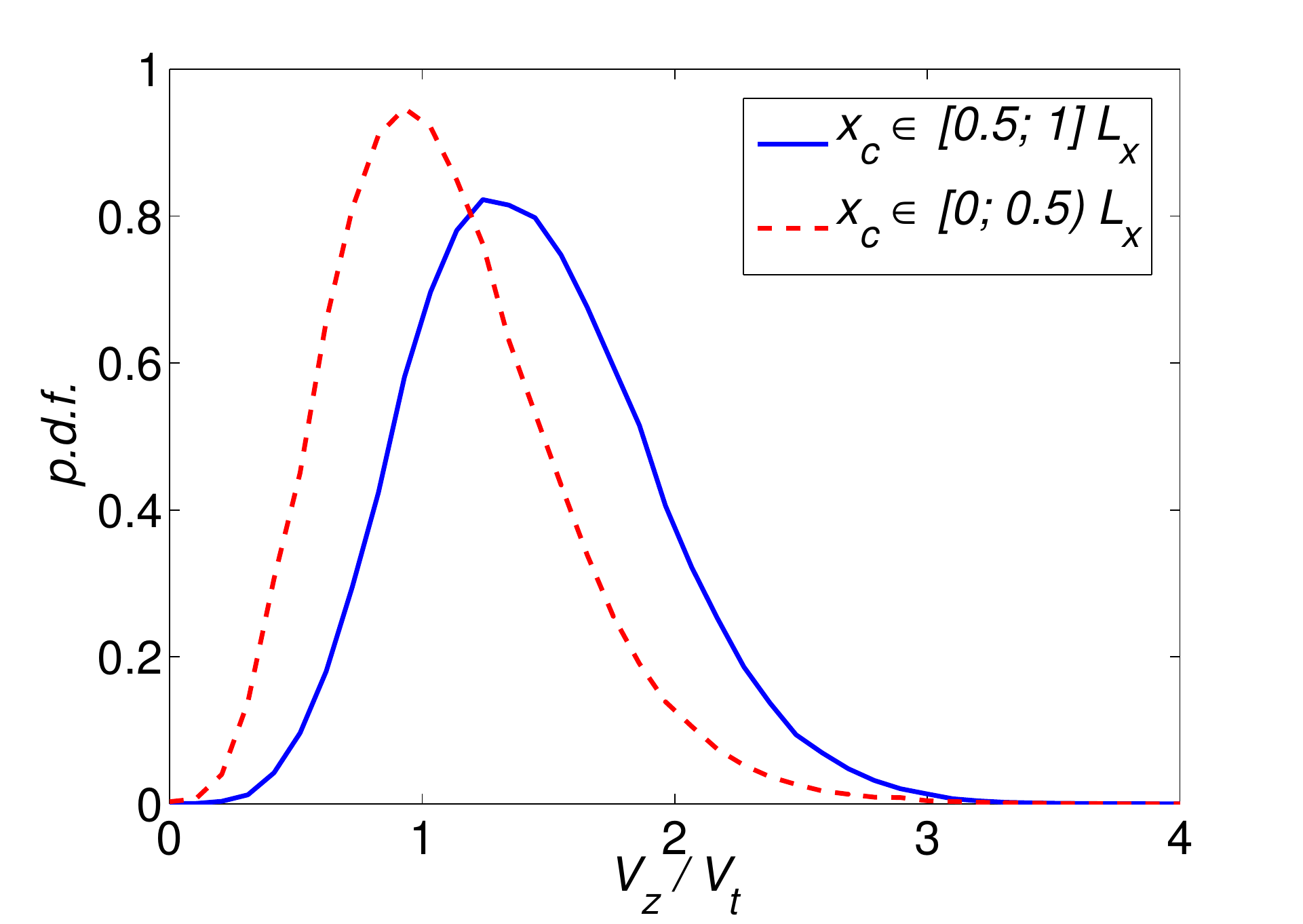}   
     \put(-194,120){{\large b)}}
     }%
\caption{(a) Probability density function, $p.d.f.$, of the settling speed for all $\phi$. (b) Probability 
density functions of the settling speed calculated for particles contained in the volumes defined by 
$x_c \in [0; 0.5) L_x$ (dashed line) and $x_c \in [0.5; 1] L_x$ (solid line), for $\phi=0.5\%$.}
\label{fig:lvel}
\end{figure}

Next, we display in figure~\ref{fig:lvel}a) the probability density function, $p.d.f.$, of the settling speed 
$V_z/V_t$ for all $\phi$ under investigation. The moments of the $p.d.f.$s are reported in table~\ref{tab:pdf}. 
For 
$\phi=0.5\%$ and $1\%$ the distributions are similar and (positively) skewed towards larger speeds than the 
mean value. As $\phi$ increases, the skeweness of the $p.d.f.$s ($S_{V_z}$) decreases becoming negligible 
for the denser case, indicating that the dynamics is mostly governed by excluded volume effects, rather 
than by pair-interactions and clustering formation. 
On the contrary, the standard deviation of the $p.d.f.$s increases with $\phi$ up to 
$\phi=5\%$. A slightly smaller $\sigma_{V_z}$ is found instead for $\phi=10\%$. 
The flatness, $F_{V_z}$, is 
always around $3$. It is interesting to observe that as $\phi$ increases, the $p.d.f.$s tend progressively 
towards a normal distribution indicating that the settling dynamics, initially governed mostly by particle 
interaction through wakes, becomes progressively dominated by the hindrance effect.

As can be seen from the $p.d.f.$s, the probability of having particles rising increases with $\phi$. This is 
due to the imposition of the zero total volume flux condition. This ensures that $\langle W_z \rangle = 
(1-\phi) \langle U_z \rangle + \phi \langle V_z \rangle = 0$, where $W_z$ and $U_z$ are the bulk and fluid 
velocities \citep{guazzelli2011}. As said, particle clusters settle substantially faster than the whole 
suspension and to satisfy the condition $\langle W_z \rangle=0$, strong upward local fluid streams are 
generated in their surroundings. When these updrafts encounter slowly settling particles, they drag them in 
the opposite direction with respect to gravity. Notice that we have observed this effect also for dense suspensions 
of spheres with $Ga \sim 9$, for which we found results in agreement with the corrected Richardson-Zaki fit 
 and with \citet{yin2007}.

In figure~\ref{fig:lvel}b) we show the $p.d.f.$s of settling speeds for particles whose centers are 
located within $x \in [0; 0.5) L_x$ or within $x \in [0.5; 1] L_x$ (i.e.\ the computational domain is divided 
in two parts denoted as left and right). Particles located within $x \in [0.5; 1] L_x$ (i.e.\ where the columnar 
structure is found, right side) settle with a mean velocity larger than that of the suspension ($\langle V_{r,z} \rangle/V_t 
= 1.45$). A smaller mean settling speed is found instead in the left half ($\langle V_{l,z} \rangle/V_t = 1.1$). 
Concerning the distribution standard deviation, this is also slightly larger in the right half, $x_c \in [0.5; 1] L_x$ (
$\sigma_r=0.48$ and $\sigma_l=0.45$). On the other hand, it is interesting to note that the skewness is 
larger for the slower particles, located in the region $x_c \in [0; 0.5) L_x$ ($S_l=0.75$ and $S_r=0.44$). This is because on the left half, 
there are less particles that less frequently undergo intense drafting-kissing-tumbling interactions. These 
interactions lead to the large skewness, while the mean value is similar to $V_t$ being the particles more 
isolated. Since the drafting-kissing-tumbling events are more intermittent in the left half, also the flatness 
is larger than for the velocities of the particles forming the fast falling column
 ($F_l=4$ versus $F_r=3$).

\begin{table}
  \begin{center}
\def~{\hphantom{0}}
  \begin{tabular}{ccccc}
      $\phi (\%)$  &   $\langle V_z \rangle$ & $\sigma_{V_z}$ & $S_{V_z}$ & $F_{V_z}$\\[3pt]
       0.5   &  1.33 & 0.50 & 0.47 & 3.11 \\
       1.0   &  1.29 & 0.55 & 0.40 & 3.15 \\
       5.0   &  1.08 & 0.66 & 0.14 & 2.89 \\
      10.0   &  0.87 & 0.60 & 0.01 & 3.06 \\
  \end{tabular}
  \caption{Moments of the $p.d.f.$s of settling speed for all $\phi$: mean value, $\langle V_z \rangle$; 
standard deviation, $\sigma_{V_z}$; skewness, $S_{V_z}$; flatness, $F_{V_z}$. The first two value are 
normalized by the terminal velocity $V_t$.}
  \label{tab:pdf}
  \end{center}
\end{table}

We now turn to the discussion of the microstructure of the whole suspension. To this aim we calculate the 
pair-distribution function $P(\vec r)$, the conditional probability of finding a particle at $\vec r$, given one at 
the origin. Following \citet{kulkarni2008}, this is defined as 
\begin{equation}
\label{pair1}
P(\vec r) = P(r,\theta,\psi) = \frac{H(r,\theta,\psi)}{n t_s \Delta \mbox{\textit{V}}} 
\end{equation}
where $\theta$ is the polar angle (measured from the positive $x$ axis), $\psi$ is the azimuthal angle (measured 
from the positive $z$ axis), $n$ is the average particle number density, $t_s$ is the total number of sampling 
points, $\Delta \mbox{\textit{V}}=r^2 \Delta r sin \psi \Delta \psi \Delta \theta$ is the volume of the sampling 
bin, and $H(r,\theta,\psi)$ is the histogram of particle-pairs. More specifically, the pair space is discretized 
in $(r,\theta,\psi)$ and at each sampling time we obtain $N_p(N_p-1)/2$ pair separation vectors $\vec r$ from the 
simulated particle configurations. Each vector $\vec r$ is put into the corresponding bin of size $\Delta V$ so that the 
particle-pair histogram $H(r,\theta,\psi)$ is progressively built.

\begin{figure}
  \centering
  \subfigure{%
    \includegraphics[scale=0.34]{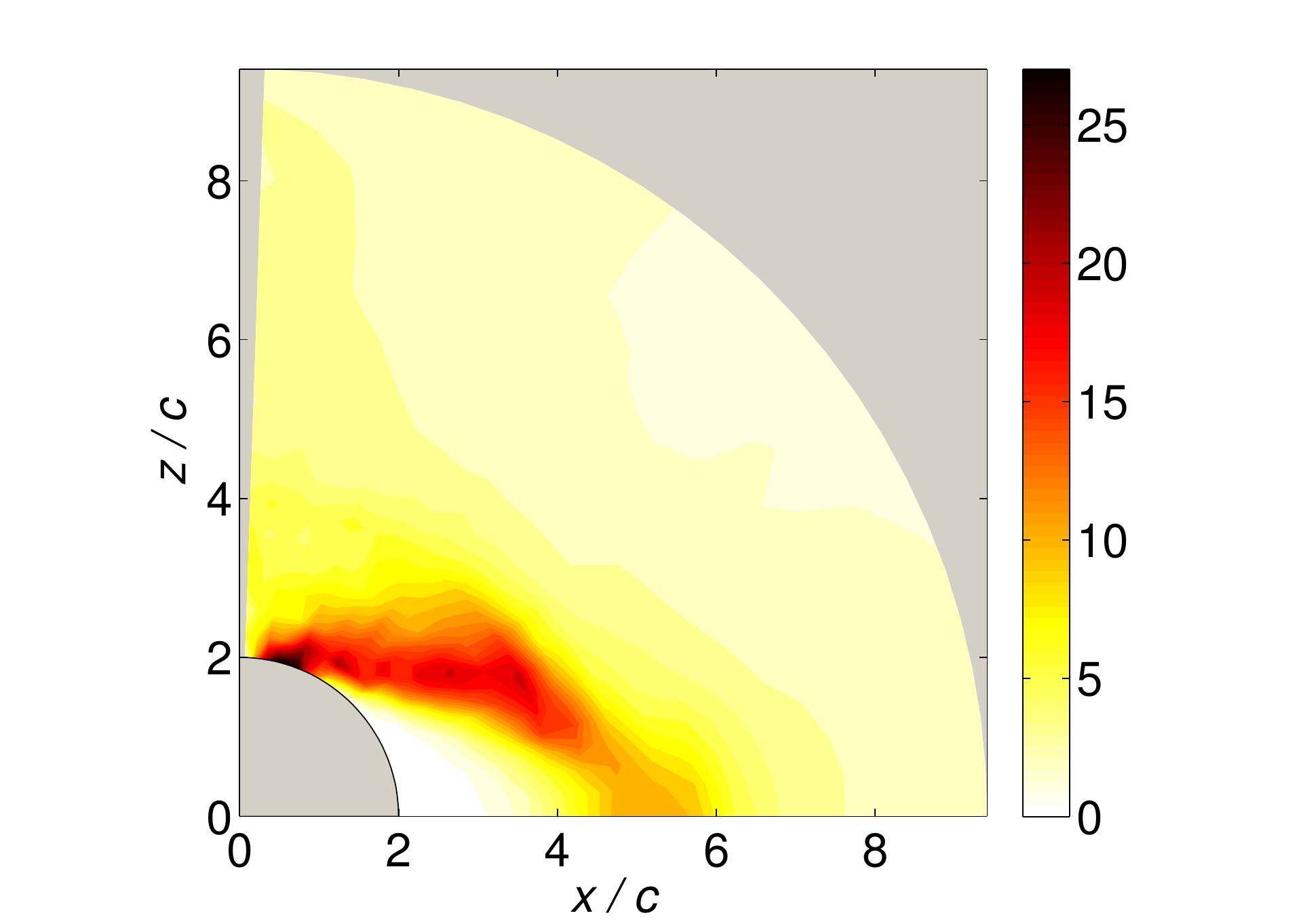}   
     \put(-184,120){{\large a)}}
     }%
  \subfigure{%
    \includegraphics[scale=0.34]{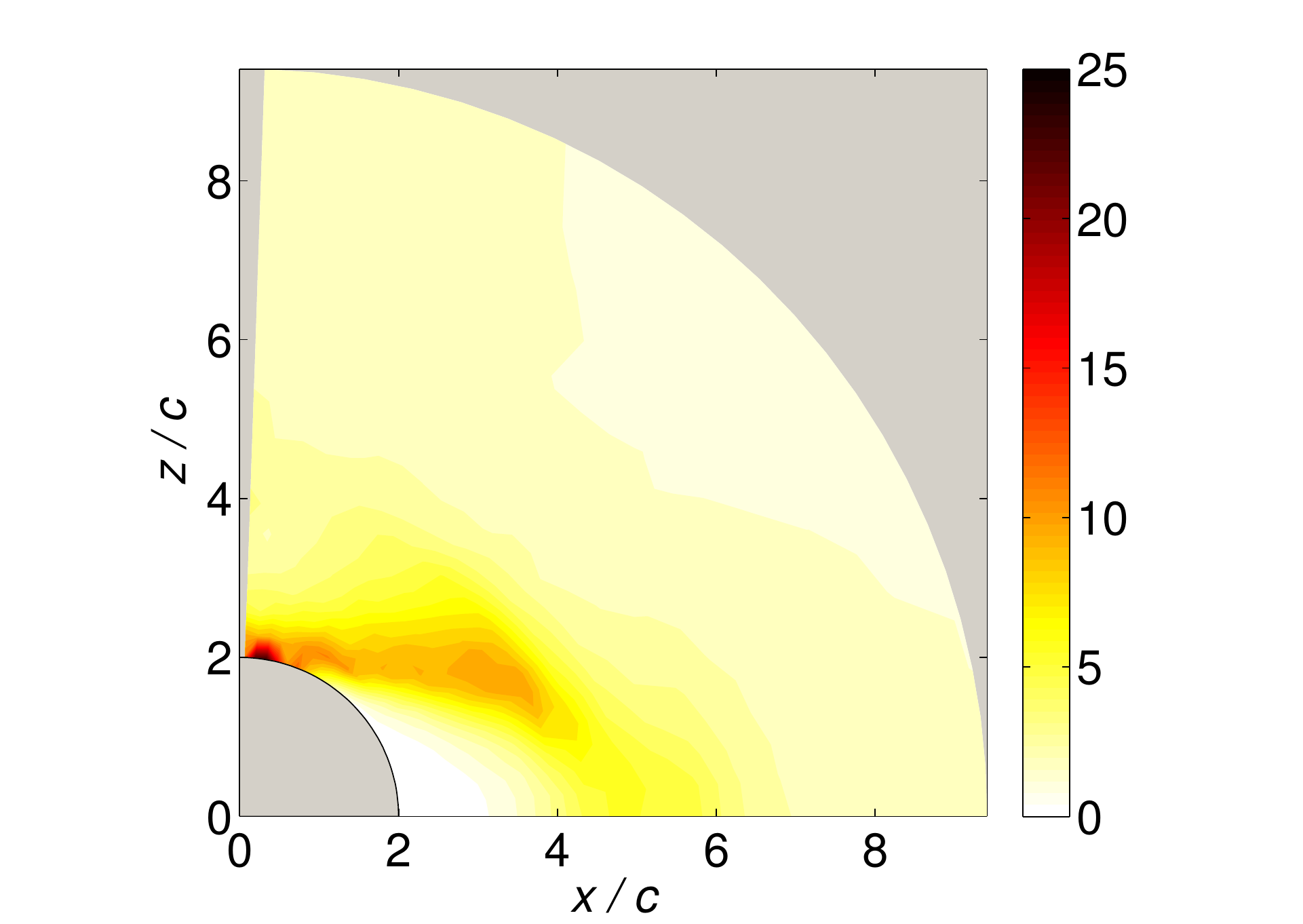}   
     \put(-184,120){{\large b)}}
     }\\%
  \subfigure{%
    \includegraphics[scale=0.34]{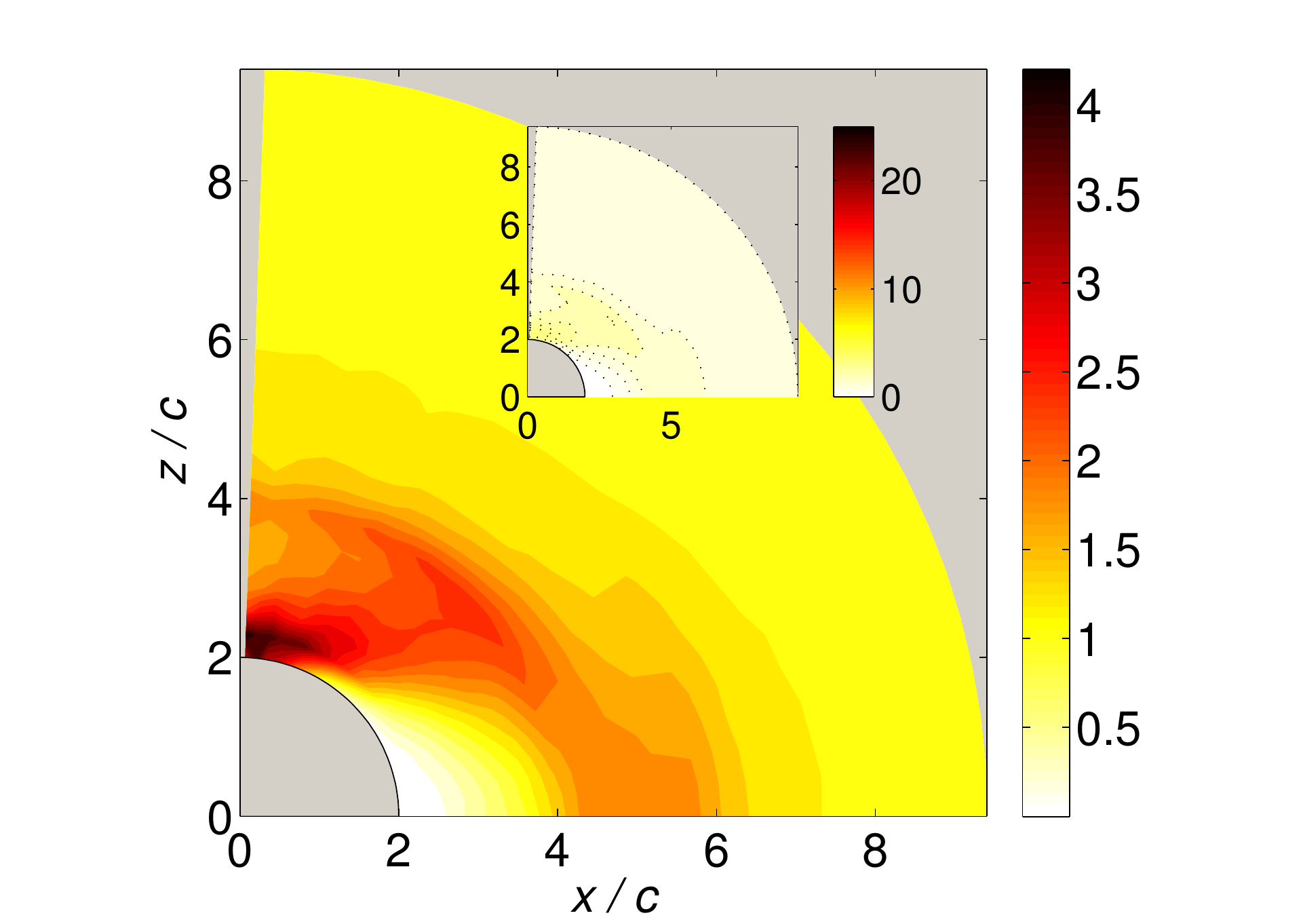}   
     \put(-184,120){{\large c)}}
     }%
  \subfigure{%
    \includegraphics[scale=0.34]{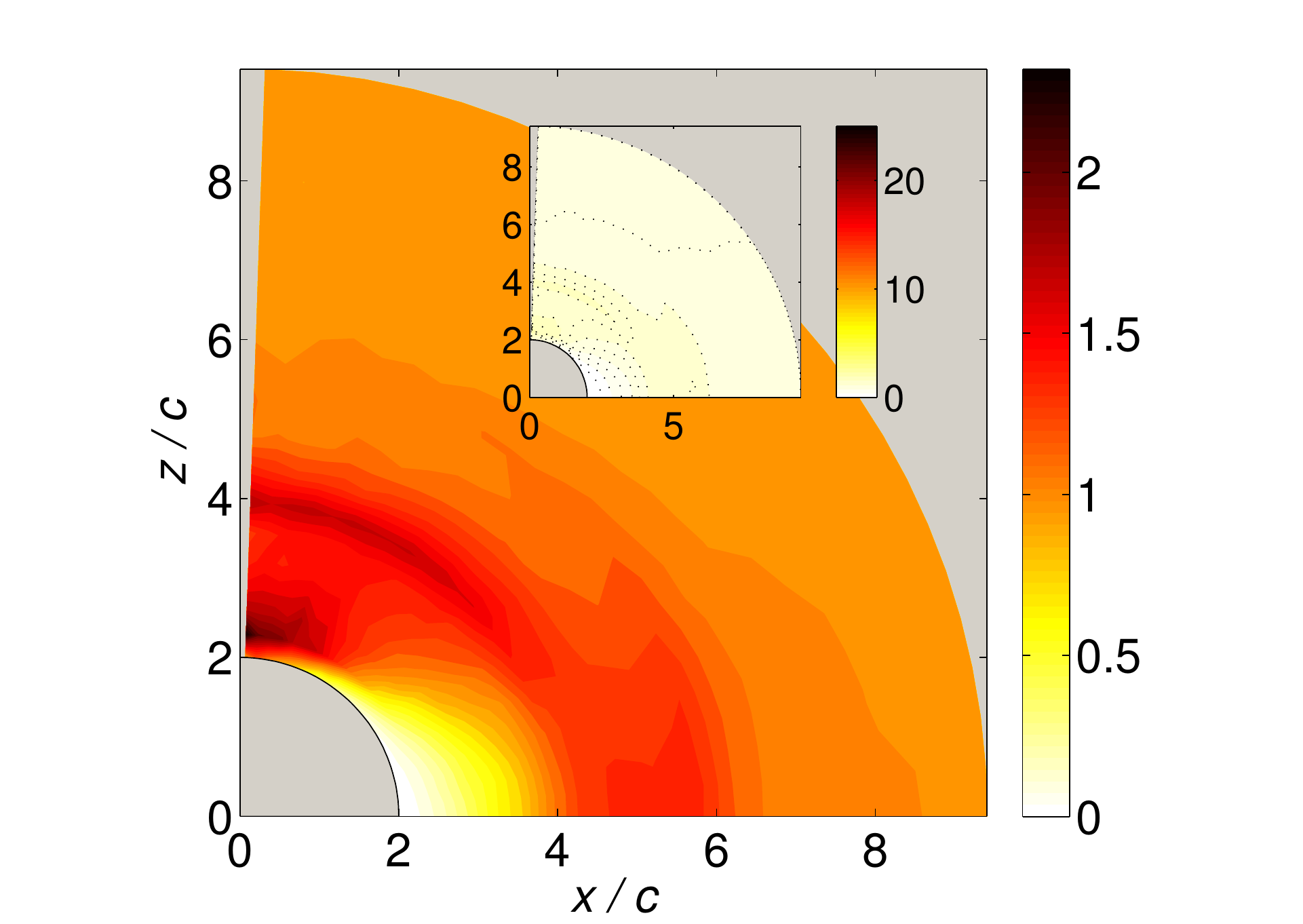}   
     \put(-184,120){{\large d)}}
     }\\%
  \subfigure{%
    \includegraphics[scale=0.34]{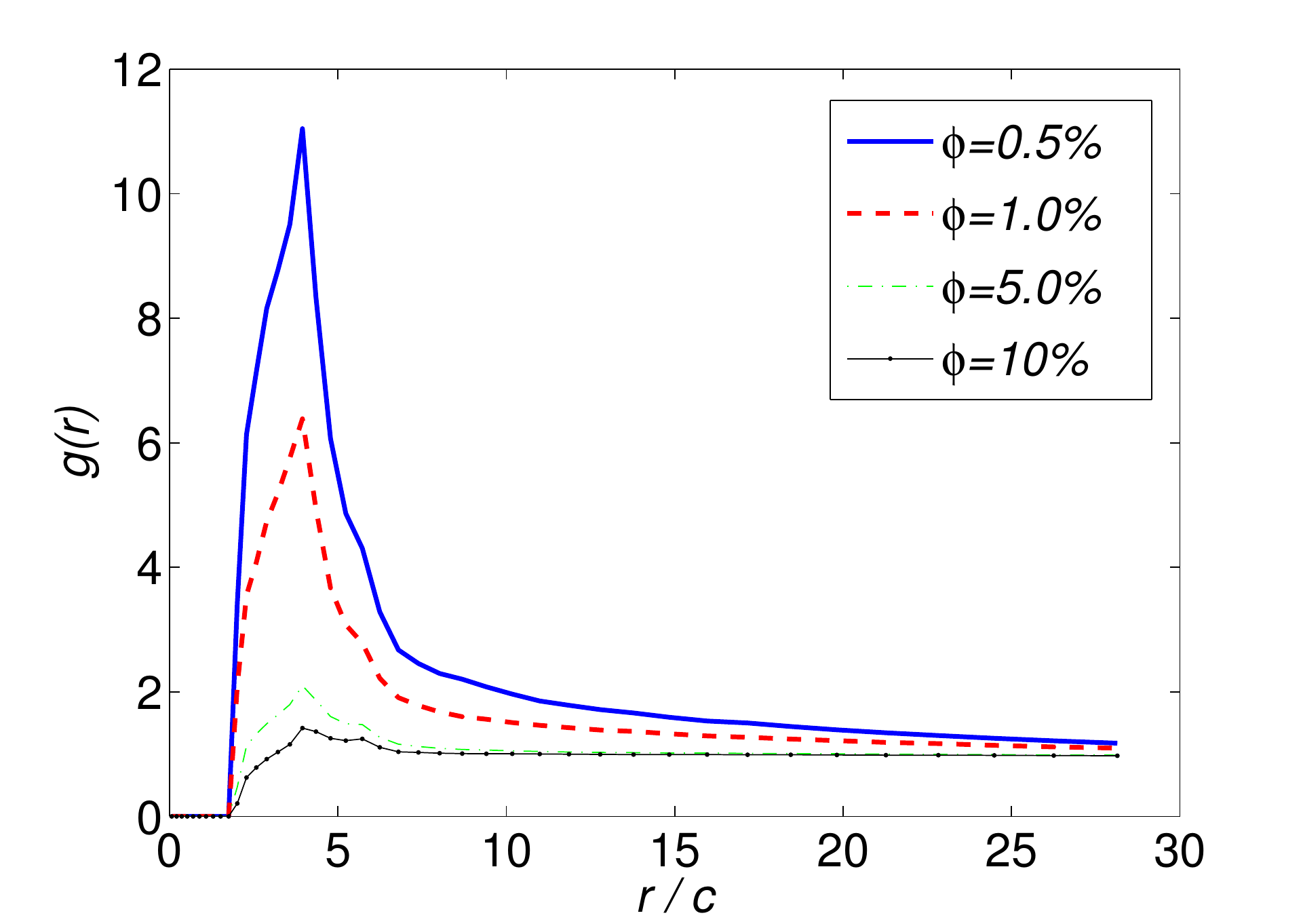}   
     \put(-184,120){{\large e)}}
     }%
  \subfigure{%
    \includegraphics[scale=0.34]{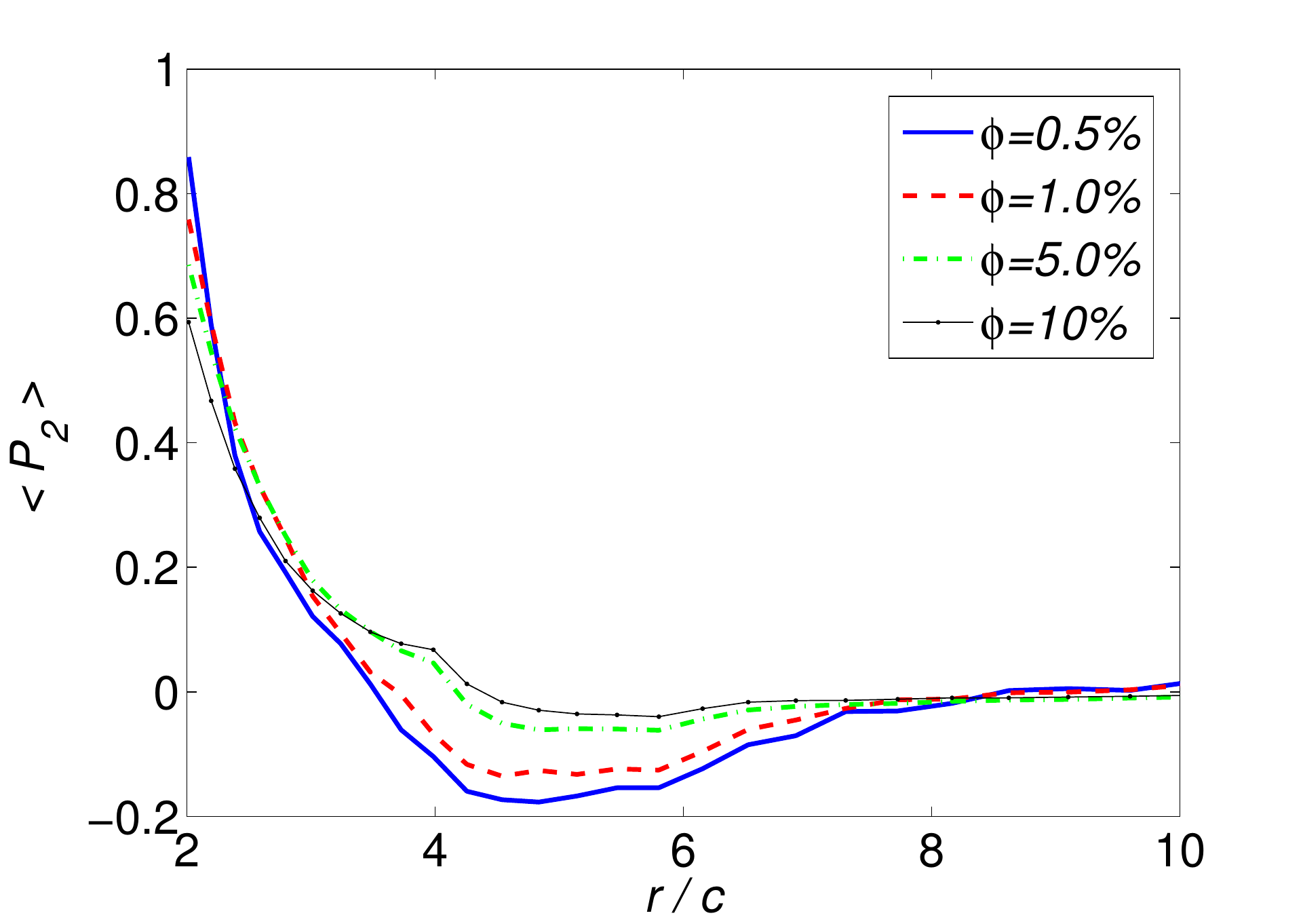}   
     \put(-184,120){{\large f)}}
     }\\%
\caption{Pair-distribution function $P(r,\theta)$ for all $\phi$ under investigation with $\phi$ increasing from left to right and top to bottom.  The insets in panels (c) and (d), corresponding to $\phi=5\%$ and $10\%$, 
show $P(r,\theta)$ with the same contour levels as in (a) and (b). Panel (e) shows the radial distribution 
function $g(r)$ for all $\phi$, while panel (f) shows the order parameter $\langle P_2 \rangle(r)$.}
\label{fig:rad}
\end{figure}

For our cases, $P(\vec r)$ is axisymmetric about the direction of gravity. Therefore we first report $P(\vec r)$ 
as function of the center-to-center distance $r$ (normalized by $c$), and the azimuthal angle $\psi$, averaging over
$\theta$. The pair-distribution function $P(\vec r)$ is shown in figure~\ref{fig:rad}a),b),c),d) for the four volume fractions under investigation. 
For simplicity, we denote the axis at $\psi=90^o$ as $x/c$. The results are substantially different from what 
observed for spheres at comparable and smaller $Ga$ \citep{yin2007,uhlmann2014} for which no evident clustering 
is observed. For $\phi=0.5\%$ we see that the pair-distribution function $P(\vec r)$ is large all around the 
reference particle. On average, each particle is surrounded by other particles, with a higher 
probability in the region between $\theta \sim 5^o$ and $\theta \sim 80^o$. From this figure it is also clear 
that particles preferentially cluster on top of each other, with their broad sides almost perpendicular to the 
vertical direction, and with an inclination that increases with $x/c$, see also figure~\ref{fig:ivel}c). The 
maximum of $P(\vec r)$ is located at $r=2.02c$ and $\psi \simeq 17^o$. Notice that horizontal clusters of 
sticking particles falling with their broad sides perfectly parallel to gravity do not occur. Indeed we see that 
$P(\vec r)=0$ for $\psi=90^o$, below $x \simeq 4c$. For $\phi=1\%$ the results are similar although $P(\vec r)$ 
is lower over the entire $(r,\theta,\psi)$ space. The probability of finding a second particle is again large all around the reference 
particle, with the highest values between $\theta \sim 2^o$ and $\theta \sim 70^o$. Particles are hence more 
vertically aligned within clusters, with a maximum $P(\vec r)$ located at $r=2.02c$ and $\psi \simeq 10^o$ (i.e.\ with the symmetry axis almost parallel to gravity). The 
maximum value is just slightly smaller than that found for $\phi=0.5\%$. The region of high $P(\vec r)$ is found 
to translate towards $\psi = 0^o$.

For $\phi=5\%$ and $10\%$, $P(\vec r)$ decreases substantially for all values of $(r,\theta,\psi)$. The maximum of 
$P(\vec r)$ is now found between $\psi=0^o$ and $2^o$. Note also that for these cases $P(\vec r) \sim 1$ for 
$r \geq 6c$, indicating that 
the random (Poissonian) distribution of particles (i.e.\ an uncorrelated statistical particle distribution) is 
already reached above this radial distance. Conversely, for the cases with $\phi=0.5\%$ 
and $1\%$ we see that $P(\vec r) \sim 1$  only in a small region between $[5; 9]c \times [30^o; 70^o]$. 
The correlation of the particle distribution at large $r/c$ for low volume fractions is indicative of the presence of the above-mentioned 
columnar structures.

A measure of the suspension microstructure that can be more easily quantified in a plot is the average of 
$P(\vec r,\theta,\psi)$ over all possible orientations. This is known as the radial distribution function $g(r)$ and 
it is shown in figure~\ref{fig:rad}e) for all volume fractions. We see that by averaging over all $\theta$ and $\psi$, the 
maximum of the radial distribution function appears at a radial distance of approximately $4c$. The peak is almost halved as $\phi$ is doubled 
from $0.5\%$ to $1\%$. For these cases, the decorrelation of the particle distribution, $g(r) \sim 1$, occurs for $r > 20c$. 
This is also approximately the radius of the columnar structure identified in figure~\ref{fig:ivel}.
The extent of clustering is sharply reduced for $\phi \geq 5\%$. Indeed, the maxima of $P(\vec r)$ are between $2$ 
and $1.5$, and the uncorrelated value (i.e. $1$) is quickly reached, $r/c \sim 6$.

Finally, we consider in figure~\ref{fig:rad}f) the order parameter. This is used to quantify the 
preferential orientation of particle pairs and it is a function of the radial separation $r$ \citep{yin2007}.
This is defined as the angular average of the second Legendre polynomial
\begin{equation}
\label{leg}
\langle P_2 \rangle (r) = \frac{\int_0^{\pi} P(r,\psi) P_2(cos \psi) sin \psi d\psi}{\int_0^{\pi} P(r,\psi) sin \psi d\psi}
\end{equation}
where $P_2(cos \psi)=(3 cos^2 \psi - 1)/2$. The order parameter $\langle P_2 \rangle (r)$ is $1$ for vertically 
aligned pairs at a separation $r$, $-1/2$ for horizontally aligned pairs and $0$ for isotropic configurations. We 
see that for $r \sim 2c$, the order parameter $\langle P_2 \rangle$ is between $0.6$ and $0.8$ for all $\phi$. 
This confirms the observation that when the separation distance $r$ is of the order of $\sim 2c$, most particles 
are almost perfectly piled up. Above $r \sim 3.5c$, $\langle P_2 \rangle$ becomes negative for $\phi=0.5\%$ and 
$1\%$, with a minimum value of approximately $-(0.2-0.15)$ around $r=4.5c$. Hence, at these radial distance 
particle pairs tend to be more horizontally aligned, with a finite inclination or pitch angle between their axis 
of symmetry and the direction of gravity plane (as we will show later). For the higher volume fractions, $\langle P_2 \rangle$ is only 
slightly negative, confirming the disappearance of strong clustering in these cases. 
The configuration becomes more isotropic after $r=8c$ for the more dilute cases, and after 
$r=6c$ for the denser cases.

\subsection{Particle dynamics}

In the previous section we have analyzed the $p.d.f.$s of particle settling speeds for all $\phi$. We now study 
the horizontal component of the translational velocity (in the direction perpendicular to gravity), as well as the 
rotational velocities.

\begin{figure}
  \centering
  \subfigure{%
    \includegraphics[scale=0.34]{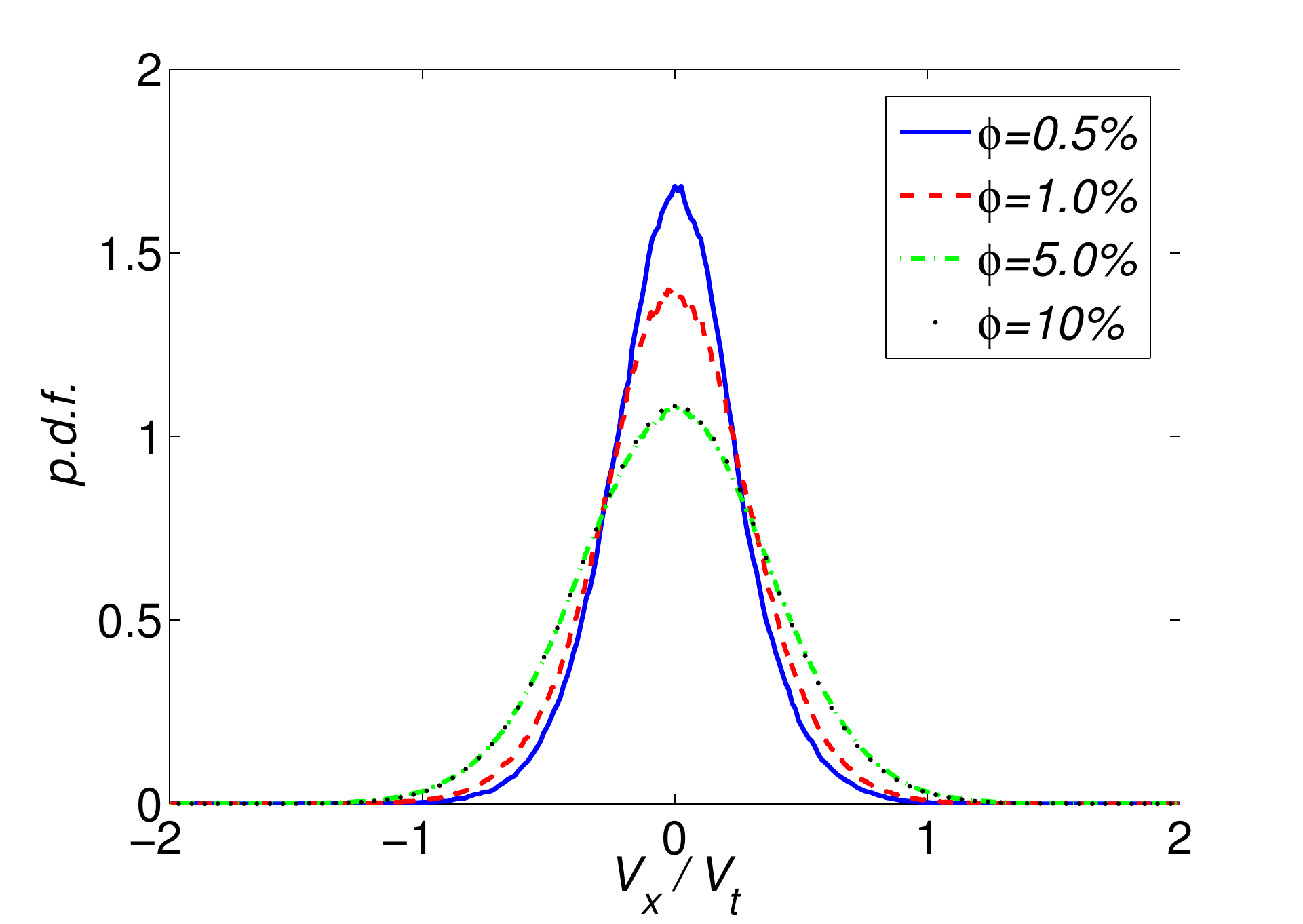}   
     \put(-188,120){{\large a)}}
     }%
  \subfigure{%
    \includegraphics[scale=0.34]{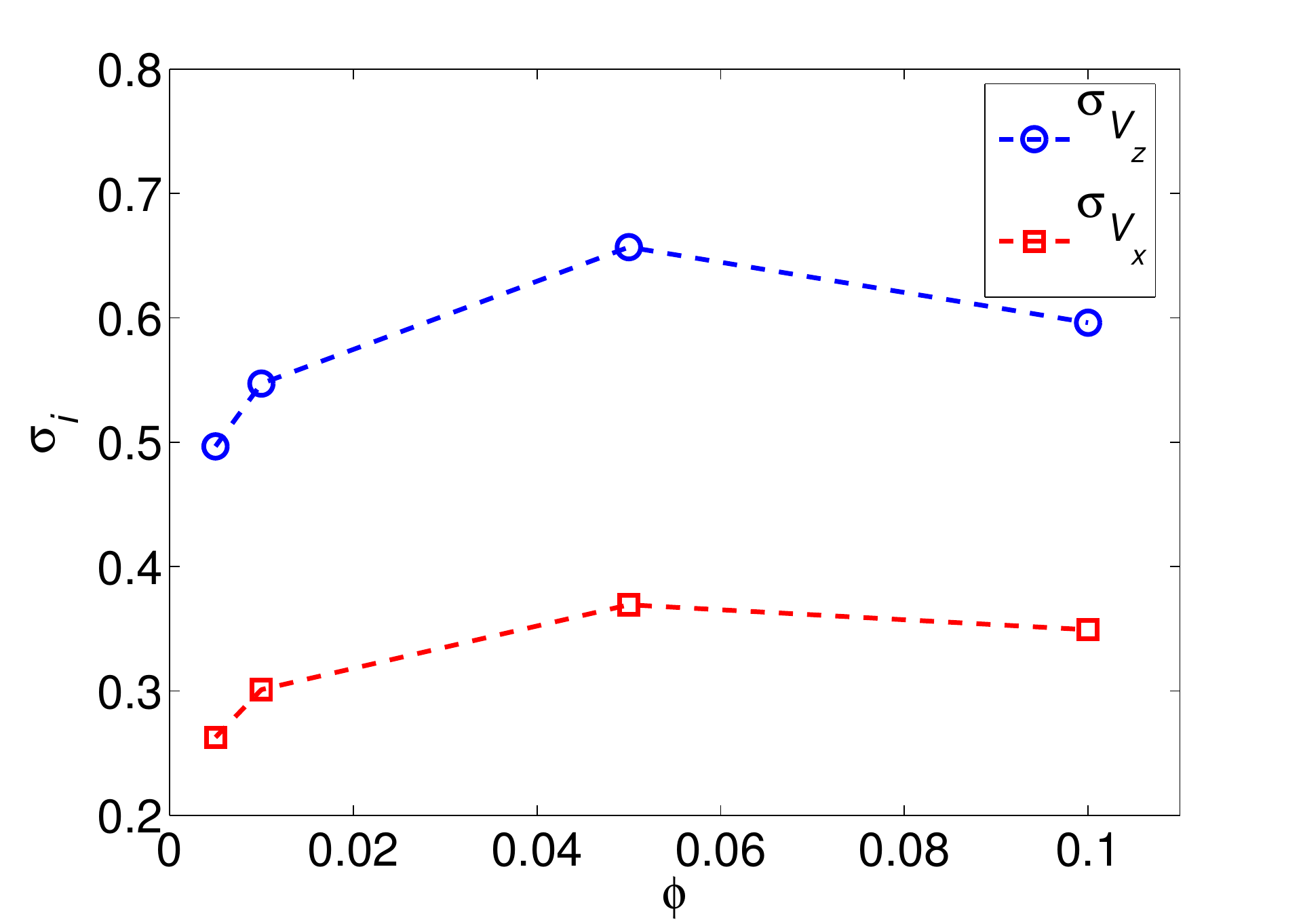}   
     \put(-188,120){{\large b)}}
     }\\%
  \subfigure{%
    \includegraphics[scale=0.34]{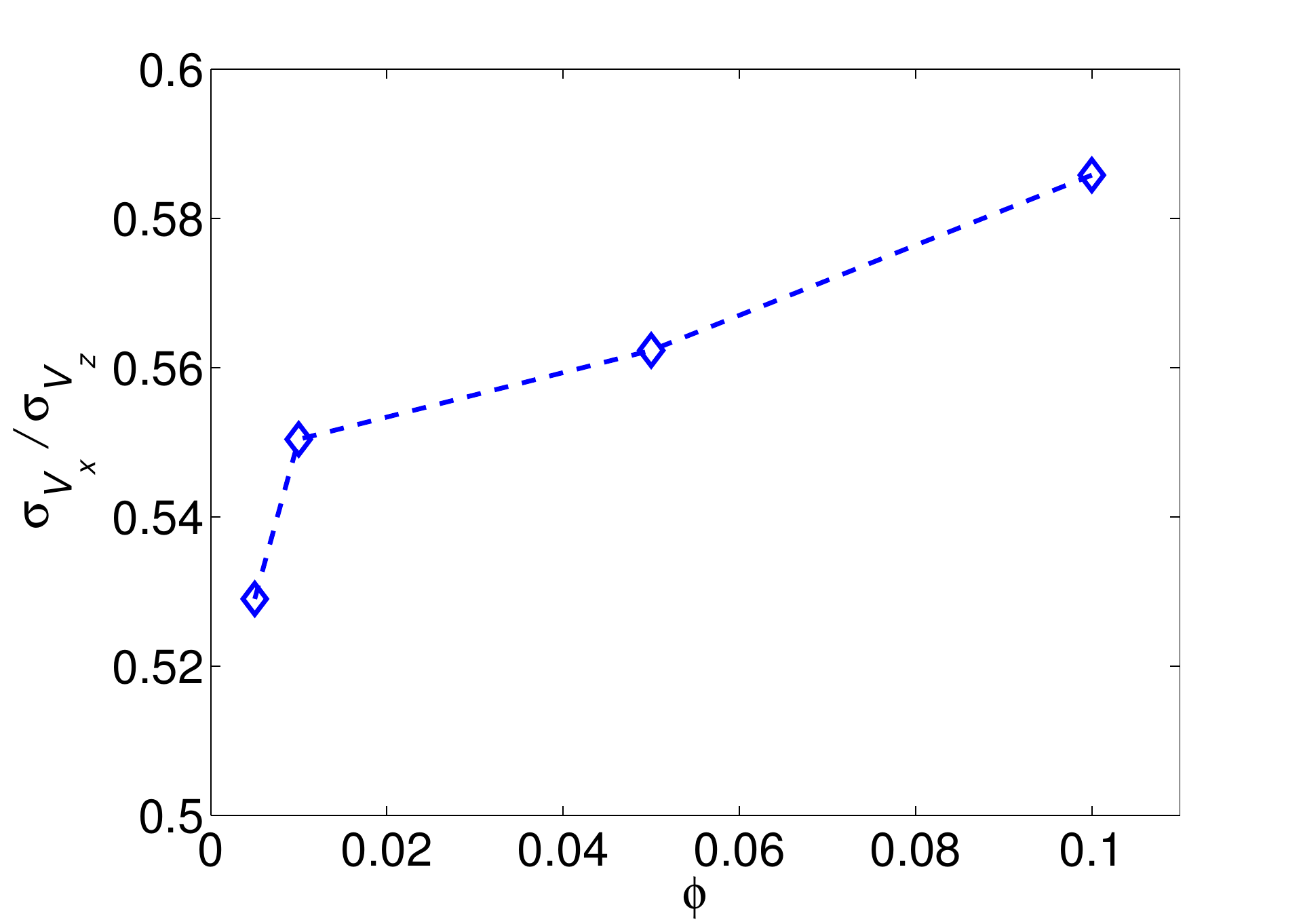}   
     \put(-188,120){{\large c)}}
     }%
  \subfigure{%
    \includegraphics[scale=0.34]{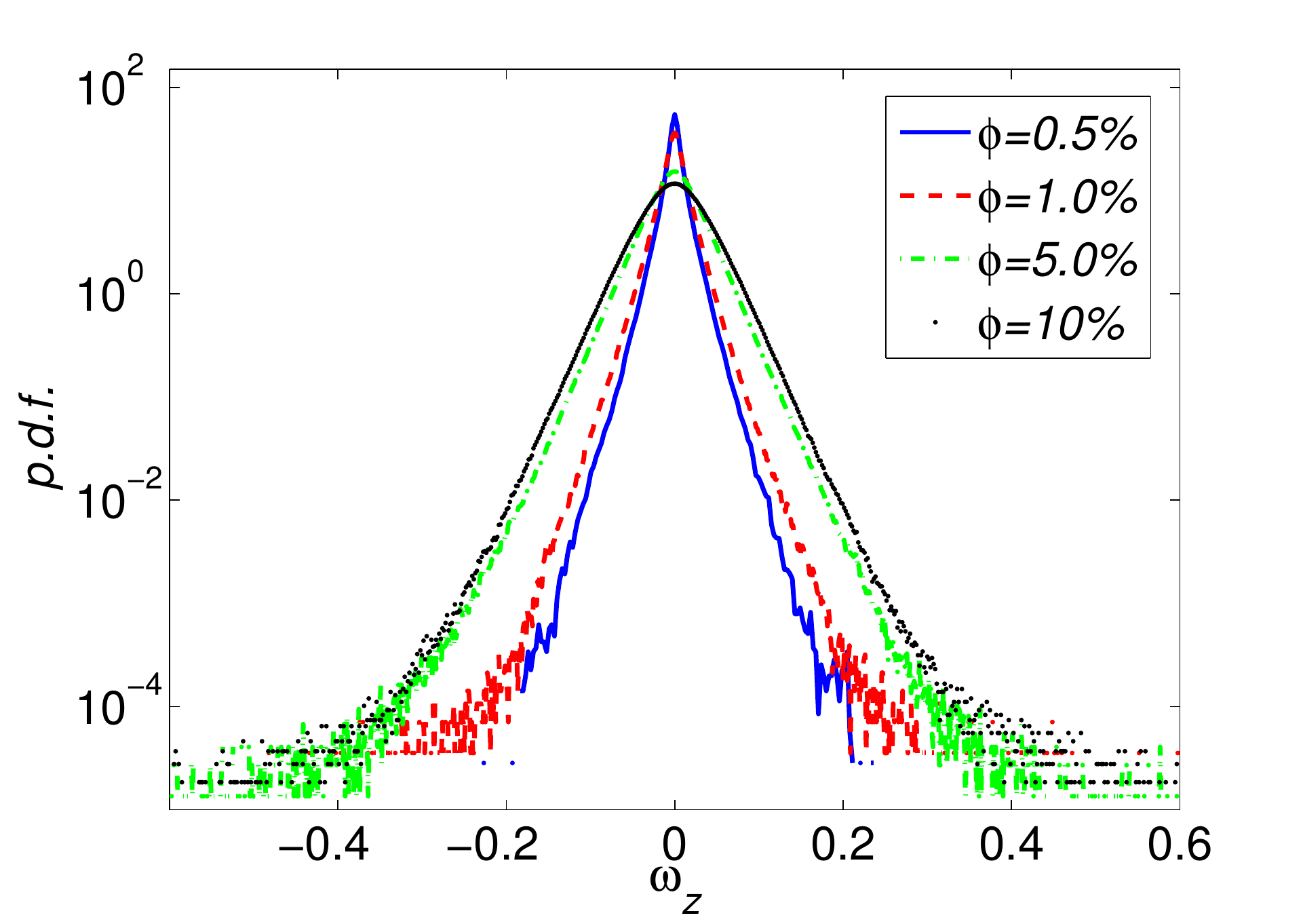}   
     \put(-188,120){{\large d)}}
     }\\%
  \subfigure{%
    \includegraphics[scale=0.34]{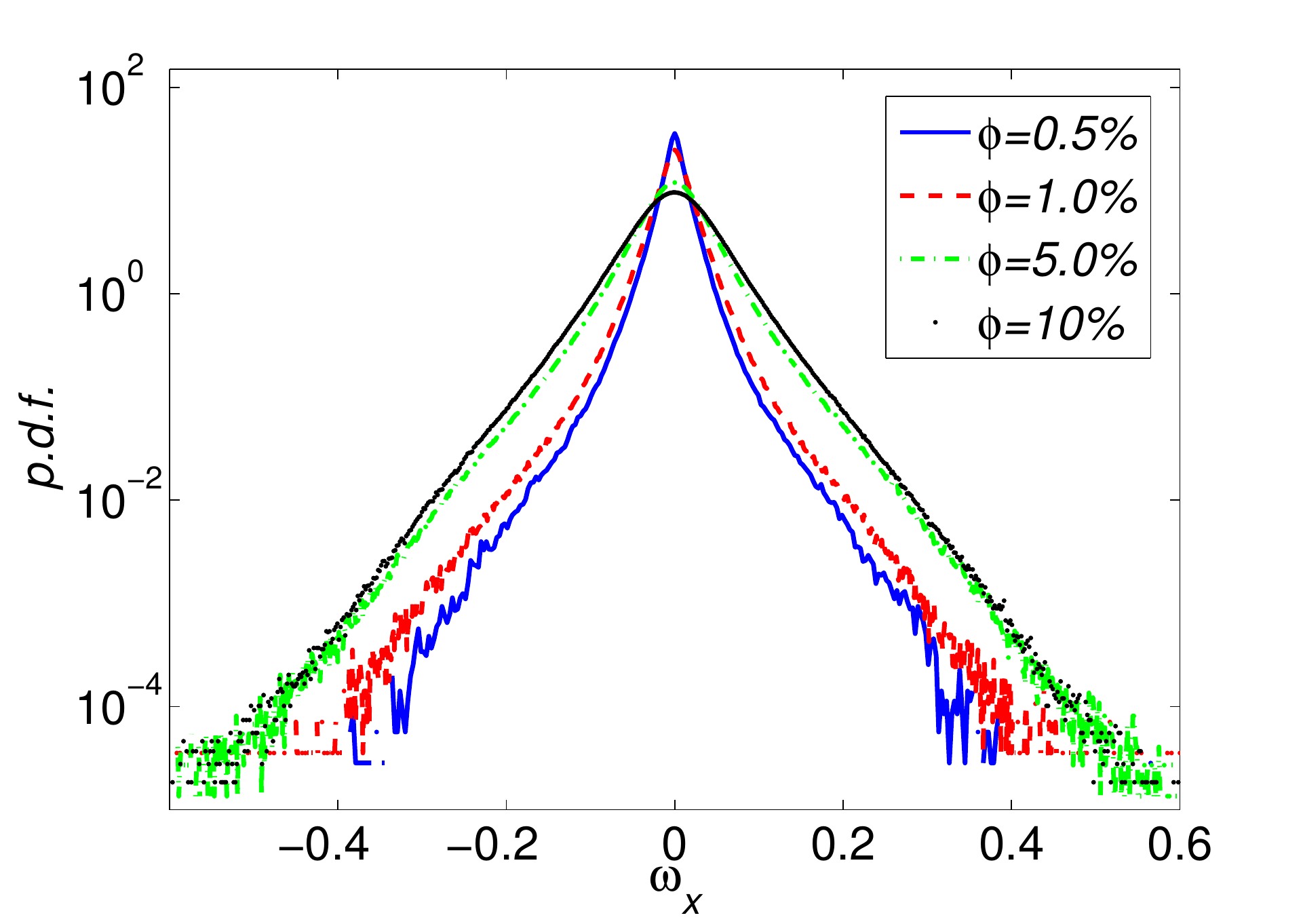}   
     \put(-188,120){{\large e)}}
     }%
  \subfigure{%
    \includegraphics[scale=0.34]{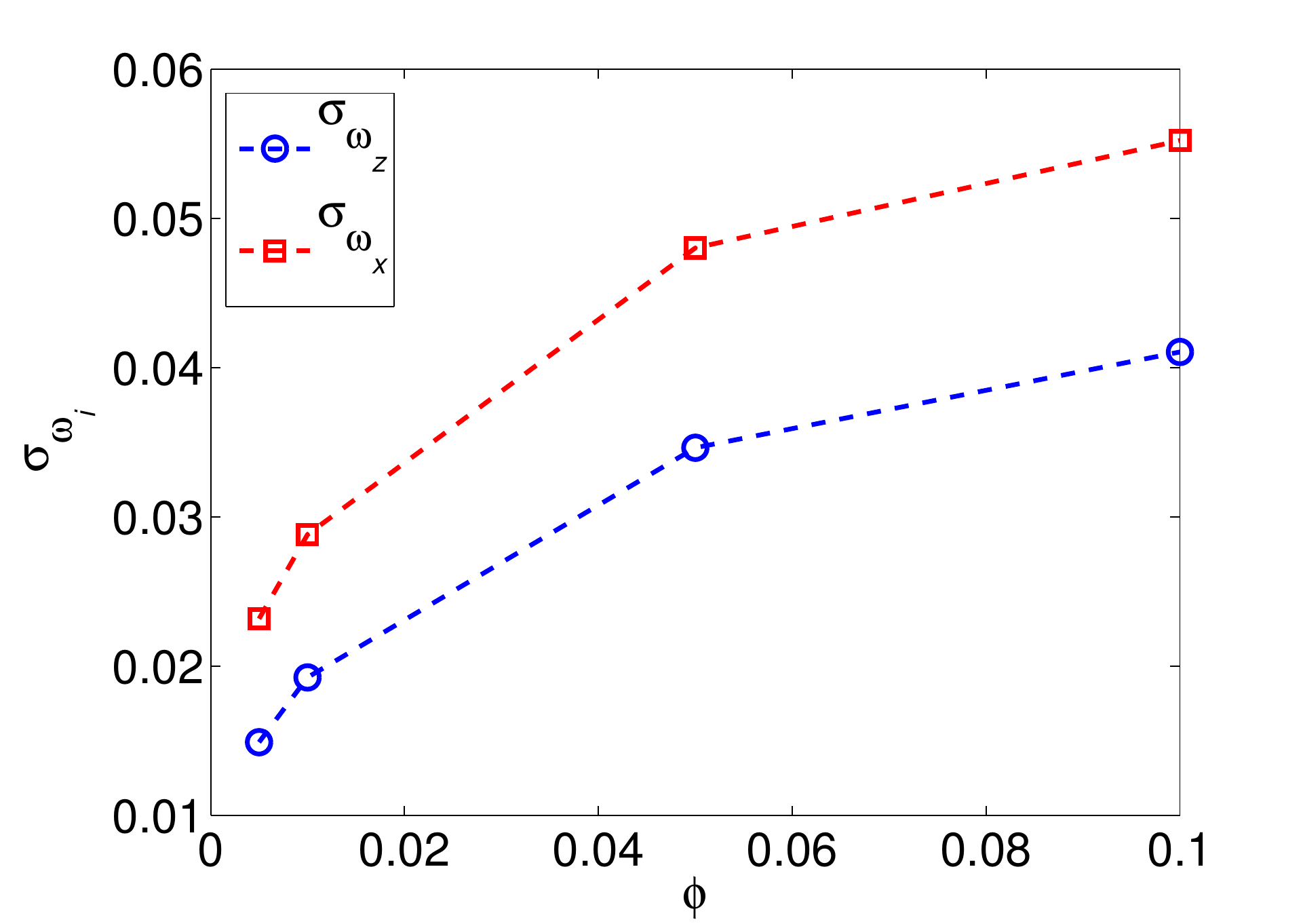}   
     \put(-188,120){{\large f)}}
     }%
\caption{(a) Particle speed in the direction perpendicular to gravity $V_{x}/V_t$ for all $\phi$. (b) Standard 
deviation of the particle velocities parallel, $\sigma_{V_z}$, and perpendicular to gravity, $\sigma_{V_x}$ as 
function of $\phi$. (c) Anisotropy of the velocity fluctuations, $\sigma_{V_x}/\sigma_{V_z}$ for all cases. (d) 
and (e), probability density functions, $p.d.f.$ of the particle angular velocities in the directions parallel, 
$\omega_z$, and perpendicular to gravity, $\omega_x$. These are normalized by $V_t/c$ ($c$ is the polar 
radius of the oblates). (f) Standard deviation of the particle angular velocities in the directions parallel and 
perpendicular to gravity.}
\label{fig:om}
\end{figure}

The probability density function of the horizontal component of the particle velocity in the plane perpendicular 
to gravity is reported in figure~\ref{fig:om}a). For the sake of simplicity, we define this component of the velocity 
as $V_x$ (note that at this stage of the simulation, most particles are found within the columnar structure and hence, the 
horizontal velocity is symmetric around the direction of gravity).
The $p.d.f.$s of $V_x/V_t$ are similar to normal distributions centered around $\langle V_x 
\rangle \sim 0$, with skewness $S \sim 0$ and flatness $F$ slighlty larger than $3$, for all $\phi$. On the other 
hand, the standard deviation $\sigma_{V_x}$ increases until $\phi=5\%$. For the more dilute cases, the fact that 
both $\langle V_x \rangle \sim 0$ and $S \sim 0$ indicate that there is a constant inflow/outflow of particles 
to/from the columnar structure.\\
The standard deviation $\sigma_{V_x}$ as function of $\phi$ is shown in figure~\ref{fig:om}b), together with 
$\sigma_{V_z}$. It is interesting that both $\sigma_{V_x}$ and $\sigma_{V_z}$ decrease after $\phi=5\%$. This is
probably an excluded volume effect as at this high $\phi$, particles are more uniformly distributed and settle as 
a bulk. Next, we display the ratio $\sigma_{V_x}/\sigma_{V_z}$ (i.e. the anisotropy of the 
velocity fluctuations), see figure~\ref{fig:om}c). 
Considering that for $1$ particle $\sigma_{V_x}/\sigma_{V_z}
=0$, we see a sharp increase of the ratio $\sigma_{V_x}/\sigma_{V_z}$ up to $1\%$. 
 Above $\phi=1\%$, the dynamics is controlled by excluded volume and hindrance effects, and the increase of the anisotropy 
with $\phi$ becomes approximately linear.

Finally, we examine  
the particle angular velocities around the 
directions parallel ($z$) and perpendicular ($x$) to gravity, see  the $p.d.f.$s of in figures~\ref{fig:om}d) and e). 
The angular 
velocities are normalized by $V_t/c$. First of all, we observe that both $p.d.f.$s are centred around 
$\langle \omega_x \rangle = 0$ and $\langle \omega_z \rangle = 0$. 
Concerning the $p.d.f.$s of $\omega_z$ we 
see that the standard deviation $\sigma_{\omega_z}$ increases substantially with $\phi$ (for $\phi=10\%$ 
$\sigma_{\omega_z}$ is almost $3$ times that found for $\phi=0.5\%$), the skewness $S$ is $\sim 0$, while the 
flatness $F$ is larger than $3$ and decreases from $10$ ($\phi=0.5\%)$ to $5.8$ ($\phi=10\%$). In the $x$ 
direction, the standard deviation $\sigma_{\omega_x}$ also increases substantially with $\phi$ 
($\sigma_{\omega_x}(\phi=0.5\%)$ is $42\%$ of $\sigma_{\omega_x}(\phi=10\%)$), $S$ is again approximately $0$ and the 
flatness $F$ decreases from $16$ to $6$. The comparison between $\sigma_{\omega_x}$ and $\sigma_{\omega_z}$ is 
shown in figure~\ref{fig:om}f). Differently from the translational velocities, we observe that the fluctuations 
of angular velocities are larger in the direction perpendicular to gravity and that these increase more rapidly 
with the volume fraction $\phi$.
We see indeed in figure~\ref{fig:ivel} that instantaneously many particles are inclined with 
respect to horizontal planes. These particles may be undergoing rotations around axes perpendicular to $g$, while 
settling with an average pitch angle. 

\begin{figure}
  \centering
  \subfigure{%
    \includegraphics[scale=0.34]{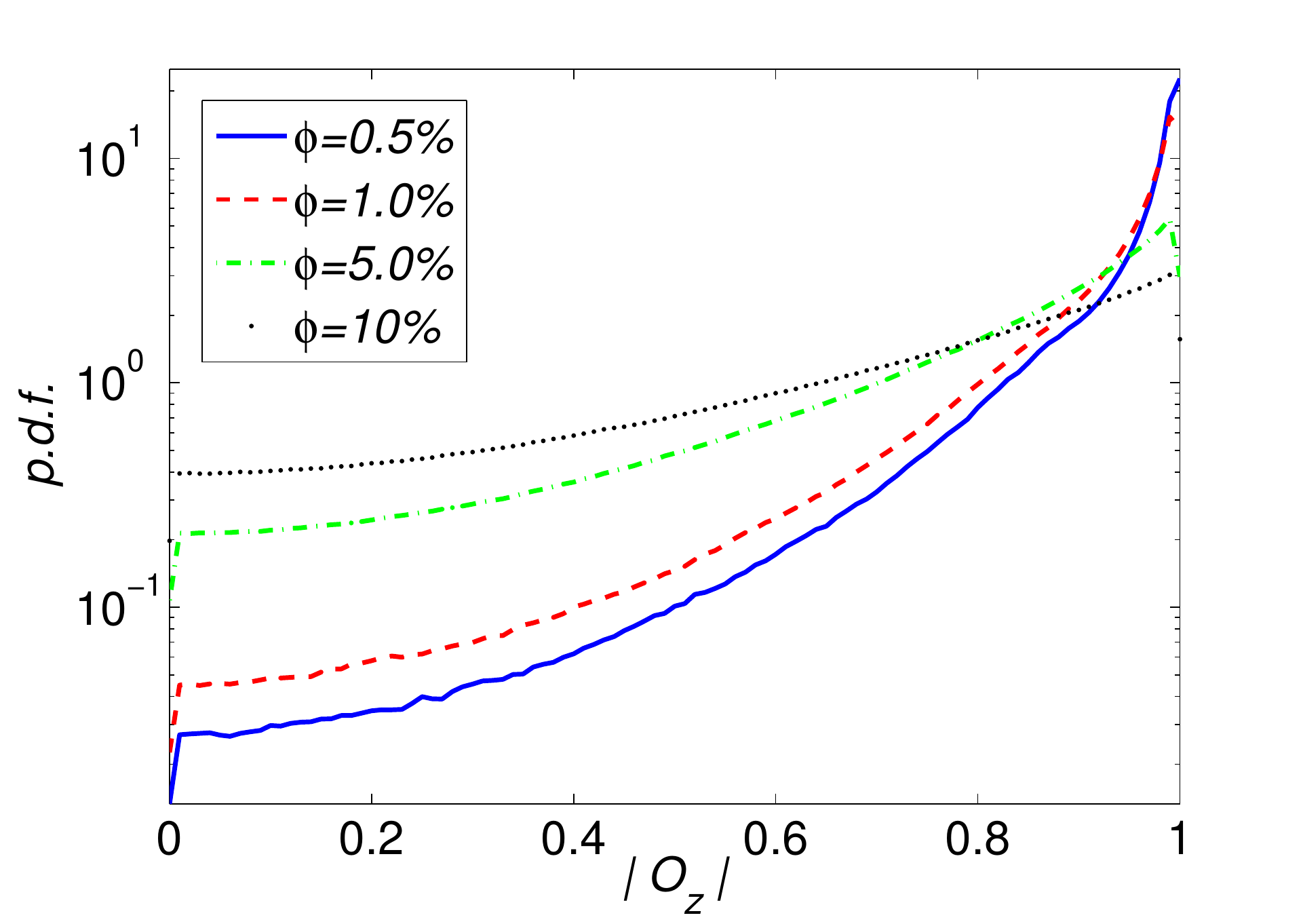}   
     \put(-188,120){{\large a)}}
     }%
  \subfigure{%
    \includegraphics[scale=0.34]{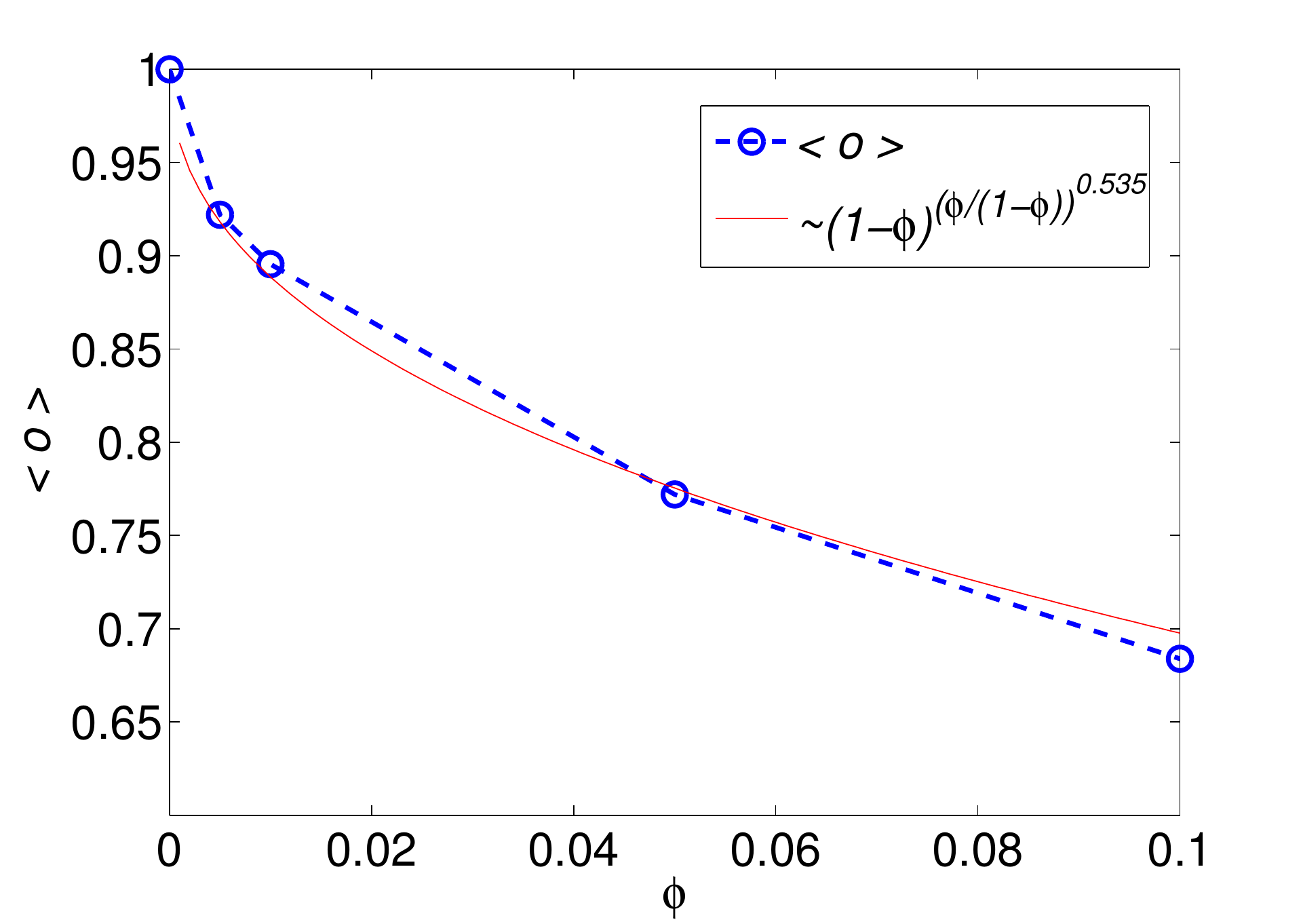}   
     \put(-188,120){{\large b)}}
     }%
\caption{(a) Probability density function of the absolute value of the particle orientation as it 
falls under gravity, $|O_z|$. (b) Mean value of the orientation $\langle |O_z| \rangle$ as function 
of $\phi$. The solid line represents an attempt to fit the data with an exponent that is itself a 
function of the volume fraction $\phi$.}
\label{fig:oz}
\end{figure}

\begin{figure}
  \centerline{\includegraphics[scale=0.2]{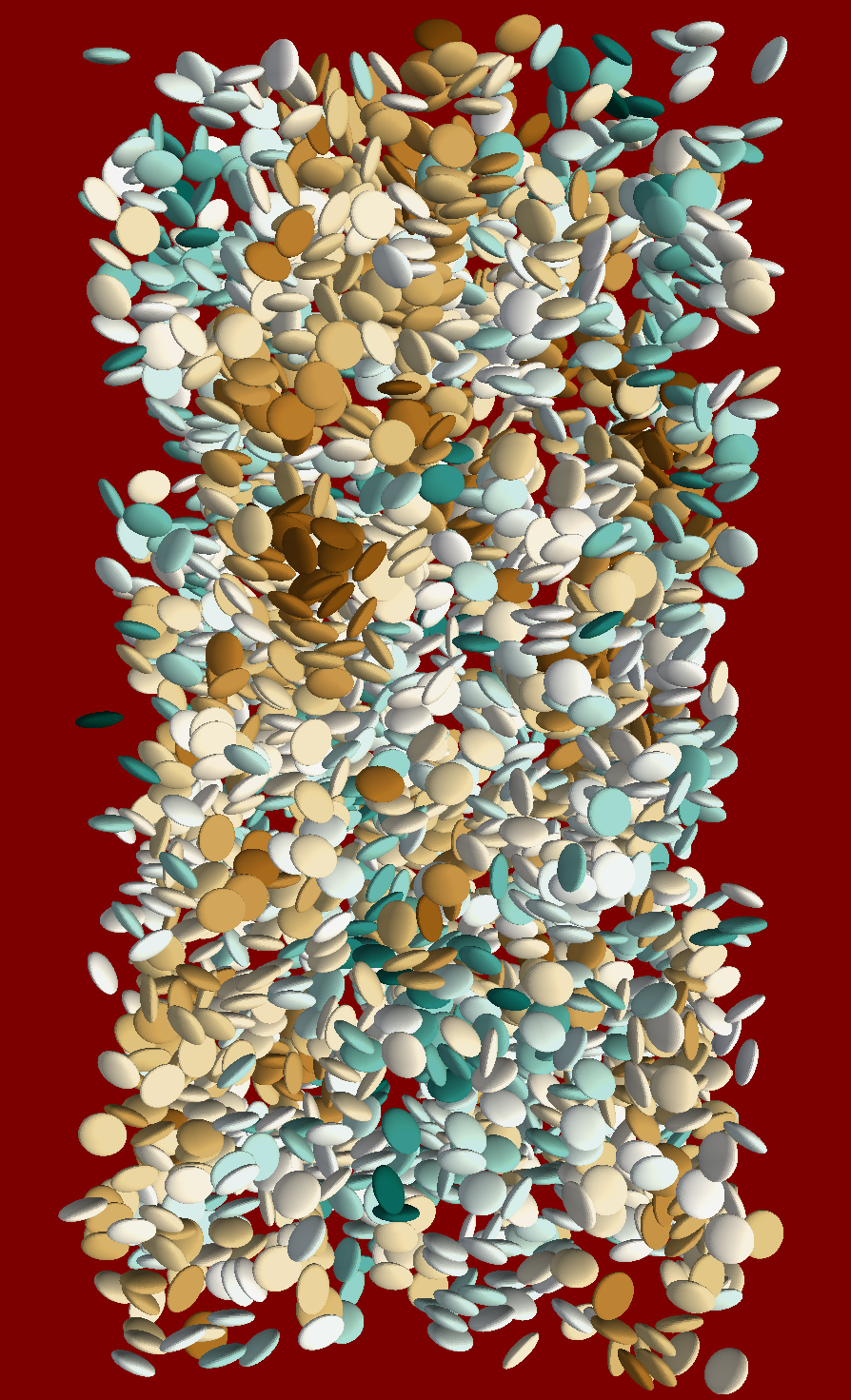}}
  \caption{Instantaneous snapshot of the suspension with $\phi=10\%$. For the sake of clarity, $25\%$ of the 
particles are shown in the first quarter of the computational domain. 
\label{fig:c100}}
\end{figure}

An isolate oblate with $Ga=60$ falls with its broad side perpendicular to the direction of gravity. The 
orientation vector, defined by the direction of the particle symmetry axis, is hence $[O_x, O_y, O_z]^T=
[0, 0, 1]$. This means that the pitch angle between the axis of symmetry and gravity is $0^o$. Due to 
hydrodynamic and particle-particle interactions in suspensions the mean particle orientation changes. 
The $p.d.f.$ of $|O_z|$ is shown for all $\phi$ in figure~\ref{fig:oz}a). 
We first observe that the probability of having particles with 
$|O_z| \simeq 0$ (i.e. with the symmetry axis aligned with gravity) increases significantly with $\phi$. For 
$\phi=10\%$ the probability of having $|O_z| \simeq 0$ is $1$ order of magnitude larger than for $\phi=0.5\%$. 
On the contrary, the mean value of $|O_z|$ decreases with $\phi$. The mean values $\langle |O_z| \rangle$ are 
shown in figure~\ref{fig:oz}b). For $\phi=0.5\%$, $\langle |O_z| \rangle=0.922$: on average particles are 
inclined by $22.8^o$ with respect to the horizontal plane. Increasing the volume fraction we find $\langle |O_z| 
\rangle=0.895$ (corresponding to an angle of $26.4^o)$, $0.765$ ($40^o$) and $0.679$ ($47^o$) for $\phi=1\%$, $5\%$, $10\%$. The increase of 
the mean pitch angle with $\phi$ is an interesting effect and indicates that particles change their orientational 
configuration to better sample the available volume. An instantaneous snapshot of the settling particles for the 
case with $\phi=10\%$ is shown in figure~\ref{fig:c100}. We see indeed that particles exhibit all possible 
orientations between $|O_z|=0$ and $1$. Note also that some particles clusters can still be observed regardless of 
the high volume fraction.

Since the change in $|O_z|$ is an excluded 
volume effect, we believe that it should be described by a function that depends directly on the volume fraction $\phi$ and on the remaining parameters, 
$Ga, R$ and $\AR$, only via some coefficients. 
Therefore we also report in figure~\ref{fig:oz}b) the function
\begin{equation}
\label{phiphi}
f(\phi) = (1-\phi)^{\left[\phi/(1-\phi)\right]^{0.535}}
\end{equation}
that is shown to fit our data sufficiently well. This observation can have implications for the modelling of settling 
suspensions.

It must be noted that if a single oblate is constrained to fall with a finite pitch angle ($O_z < 1$), it will 
reach a terminal velocity larger than that for $O_z=1$. We hence decided to perform an additional simulation of an 
isolated oblate settling with the mean pitch of the $\phi=0.5\%$ case ($O_z=0.921$ or $22.8^o$) to see how the terminal velocity 
$V_t$ of an inclined particle compares to $\langle V_z \rangle$. We find that $V_x/V_t=0.41$ and $V_z/V_t=1.03$. The increase of the 
falling speed is limited, significantly lower than that of the suspension, showing again the importance of particle-pair interactions. The drift speed is instead 
large and about $1.6 \sigma_{V_x}$ ($\phi=0.5\%$). We therefore believe that the drift speed of inclined particles within the 
suspension plays a role in the formation of the columnar structure.

\begin{figure}
  \centering
  \subfigure{%
    \includegraphics[scale=0.34]{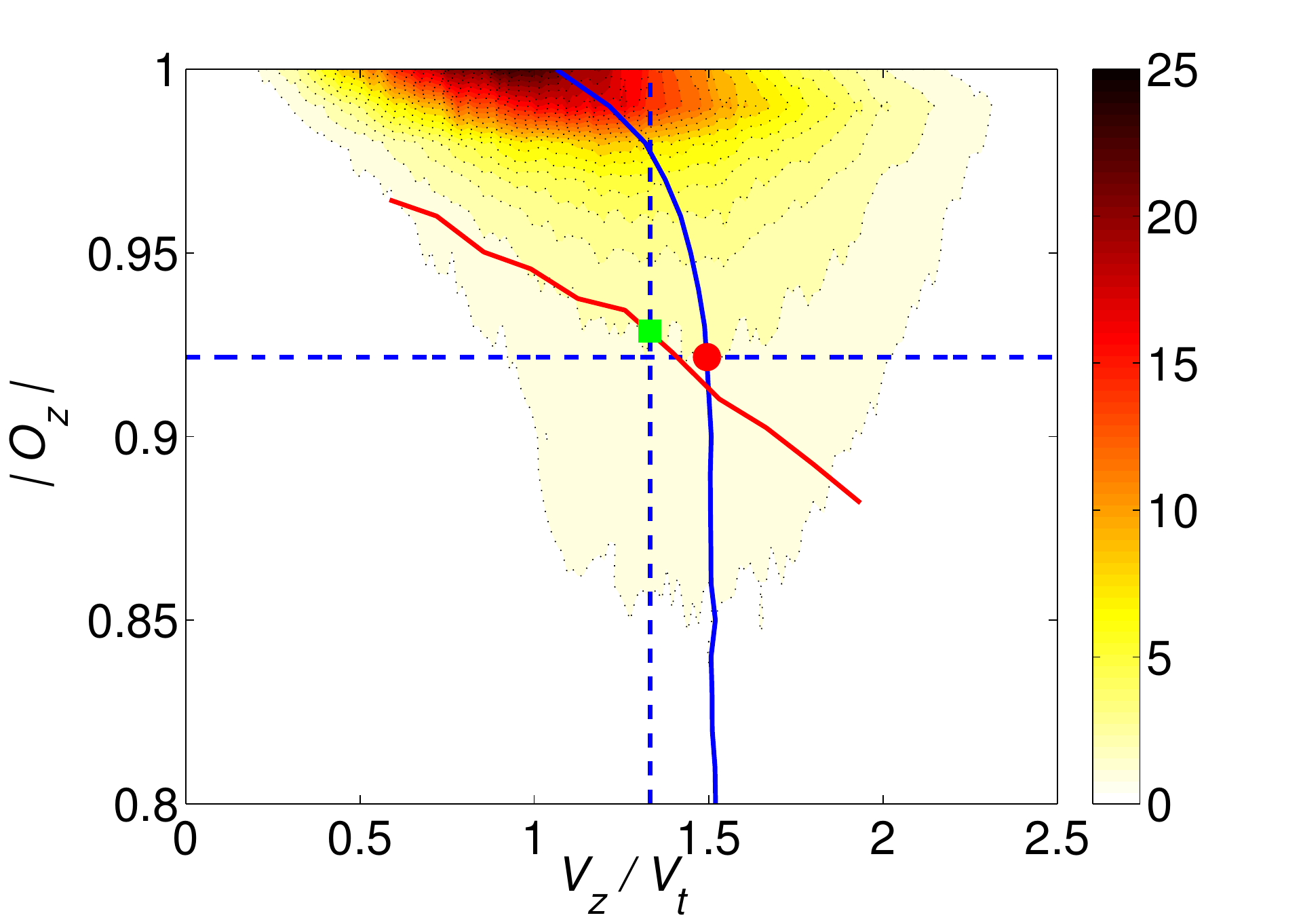}   
     \put(-188,120){{\large a)}}
     }%
  \subfigure{%
    \includegraphics[scale=0.34]{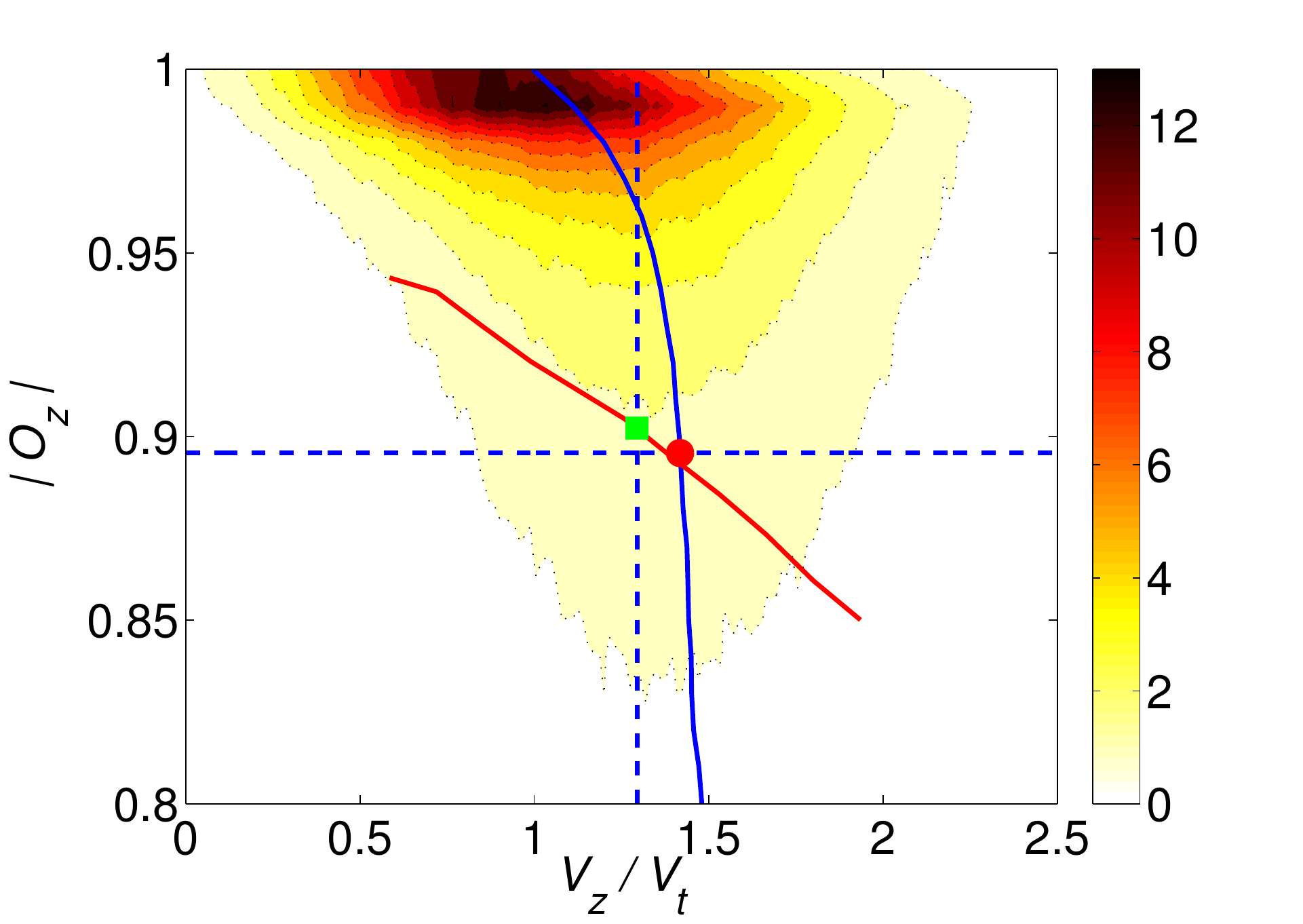}   
     \put(-188,120){{\large b)}}
     }\\%
  \subfigure{%
    \includegraphics[scale=0.34]{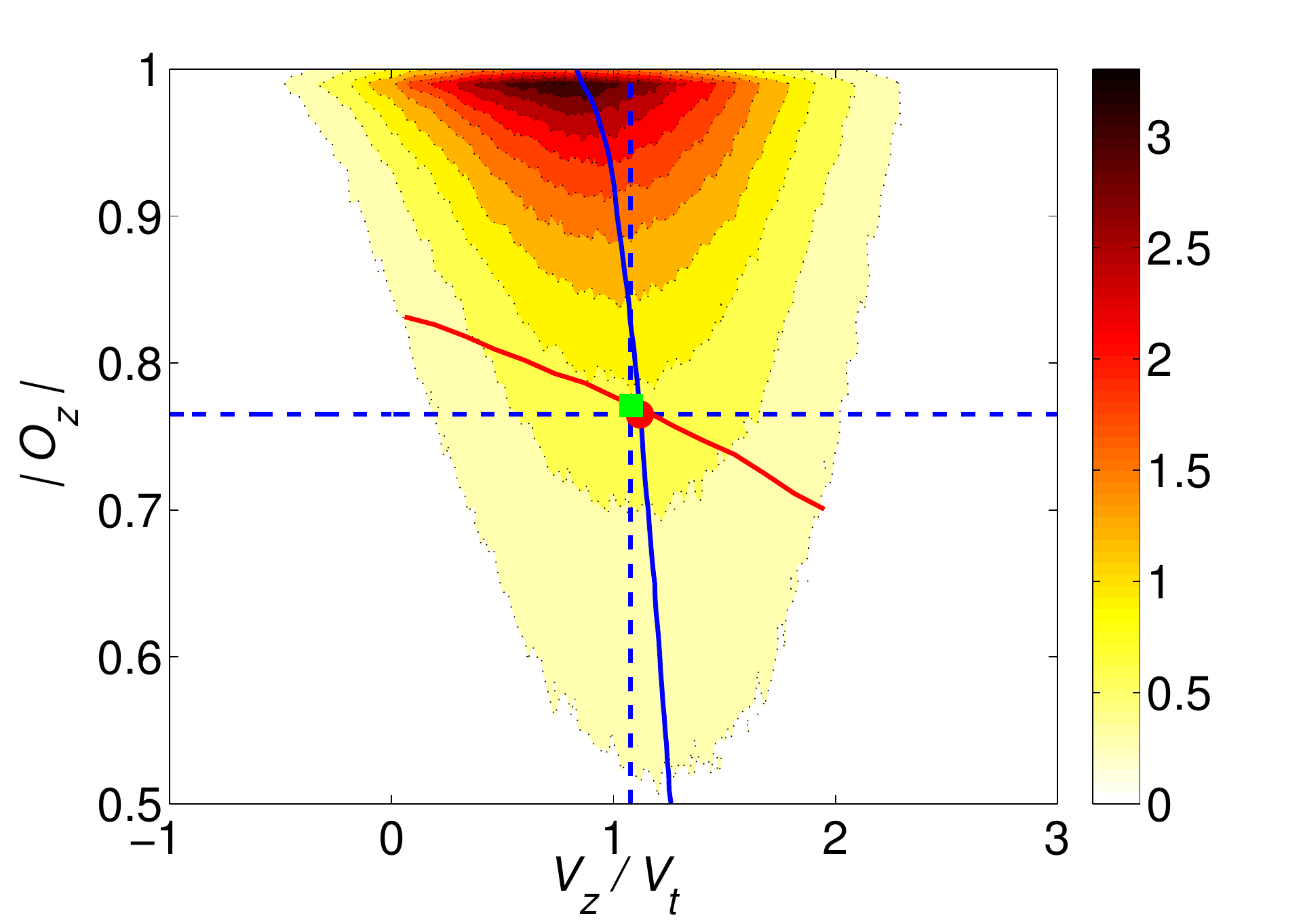}   
     \put(-188,120){{\large c)}}
     }%
  \subfigure{%
    \includegraphics[scale=0.34]{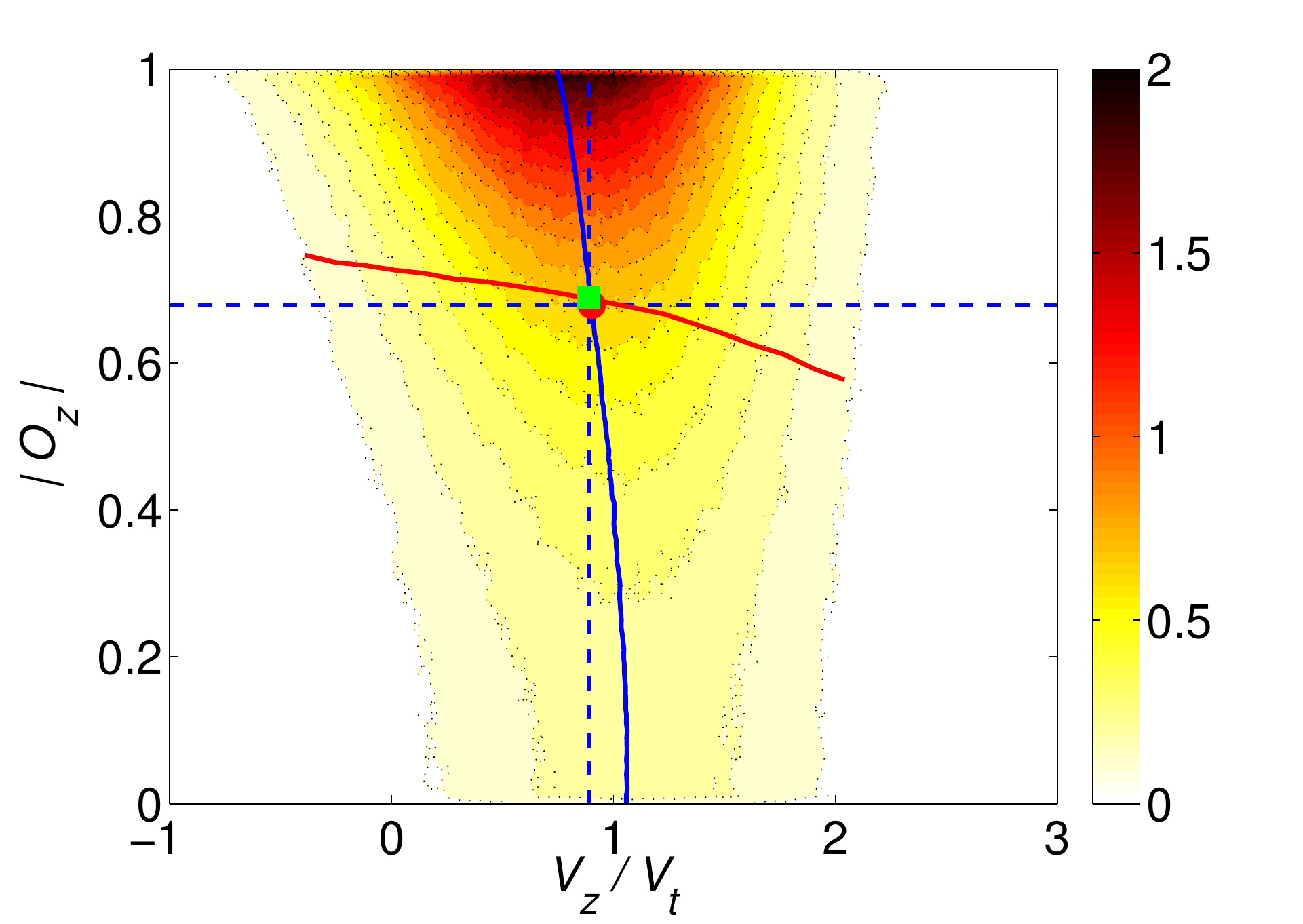}   
     \put(-188,120){{\large d)}}
     }%
\caption{Joint probabilty 
density functions of the settling speed $V_z/V_t$, and the absolute value of the orientation $|O_z|$, 
for increasing volume fraction  $\phi$ in panels (a),(b),(c) and (d). The dashed lines correspond to the mean $V_z/V_t$ and $|O_z|$. The red circles correspond 
to the mean $V_z/V_t$ conditioned to $|O_z|=\langle |O_z| \rangle$ of the suspension. The green squares 
correspond to the mean $|O_z|$ conditioned to $V_z = \langle V_z \rangle$ of the suspension. The conditioned 
averaged values for $\langle V_z/V_t \, | \, |O_z| \rangle $ and $\langle |O_z| \, | \, V_z/V_t \rangle$ are 
showed by the blue and red curves.}
\label{fig:oz1}
\end{figure}

To conclude this section we report in figures~\ref{fig:oz1}a),b),c),d) the joint probability density functions $J$ 
of particle settling speed $V_z/V_t$ and orientation $|O_z|$ for $\phi=0.5\%, 1\%, 5\%$ and $10\%$. We also show 
the mean values of $V_z/V_t$ and $|O_z|$ that we found for the suspension, indicated by dashed lines. The solid blue and red 
lines in the plots represent the mean $V_z/V_t$ and $|O_z|$ obtained by conditioned averages of $J(V_z/V_t,|O_z|)$
\begin{equation}
\label{jpdf1}
\langle V_z/V_t \, | \, |O_z| \rangle = \int_{-\infty}^{\infty} V_z J(V_z/V_t \, | \, |O_z|) \, dV_z ,
\end{equation}
\begin{equation}
\label{jpdf2}
\langle |O_z| \, | \, V_z/V_t \rangle = \int_{0}^{1} |O_z| J(|O_z| \, | \, V_z/V_t) \, d|O_z| .
\end{equation}
For $\phi=0.5\%$ and $1\%$, we find that $\langle V_z/V_t \, | \, |O_z| \rangle$ increases from about $1$ at 
$|O_z|=1$ to an almost asymptotic value of $1.5$. Hence, particles settling with an inclination $\geq acos^{-1} 
\langle |O_z| \rangle$) fall on average with speeds that are $12\%$ larger than $\langle V_z \rangle$. In 
particular, we find $\langle V_z/V_t \, | \, |O_z|=\langle |O_z| \rangle \rangle = 1.49$ and $1.42$ for $\phi=
0.5\%, 1\%$ (red circles). Concerning $\langle |O_z| \, | \, V_z/V_t \rangle$ we see that it decreases almost 
linearly with $V_z/V_t$. Particles with larger $V_z/V_t$ fall on average with smaller $\langle |O_z| \rangle$ 
(i.e. more inclined with respect to the horizontal plane). The values found at speeds equal to $\langle V_z/V_t 
\rangle$ are $1\%$ larger than $\langle |O_z| \rangle$ (green squares).\\
Similar observations apply also to the remaining cases, those at higher $\phi$. As expected, for all $|O_z|$, $\langle 
V_z/V_t \, | \, |O_z| \rangle$ decreases with $\phi$. Specifically, $\langle V_z/V_t \, | \, |O_z|=\langle |O_z| 
\rangle \rangle = 1.12$ and $0.90$ for $\phi=5\%-10\%$, about $4\%$ less than $\langle V_z \rangle/V_t$. For 
$\langle |O_z| \, | \, V_z/V_t \rangle$ we again observe values that are $1\%$ larger than $\langle |O_z| \rangle$.

Summarizing, from the joint $p.d.f.$s of particle settling speeds and falling orientation, we find that on average particles 
settling with higher velocities tend to fall with their axis more inclined with respect to the direction of gravity. Hence, as particles 
interact through their wakes, eventually forming clusters, they tend to increase their pitch angle.

\begin{figure}
  \centering
  \subfigure{%
    \includegraphics[scale=0.34]{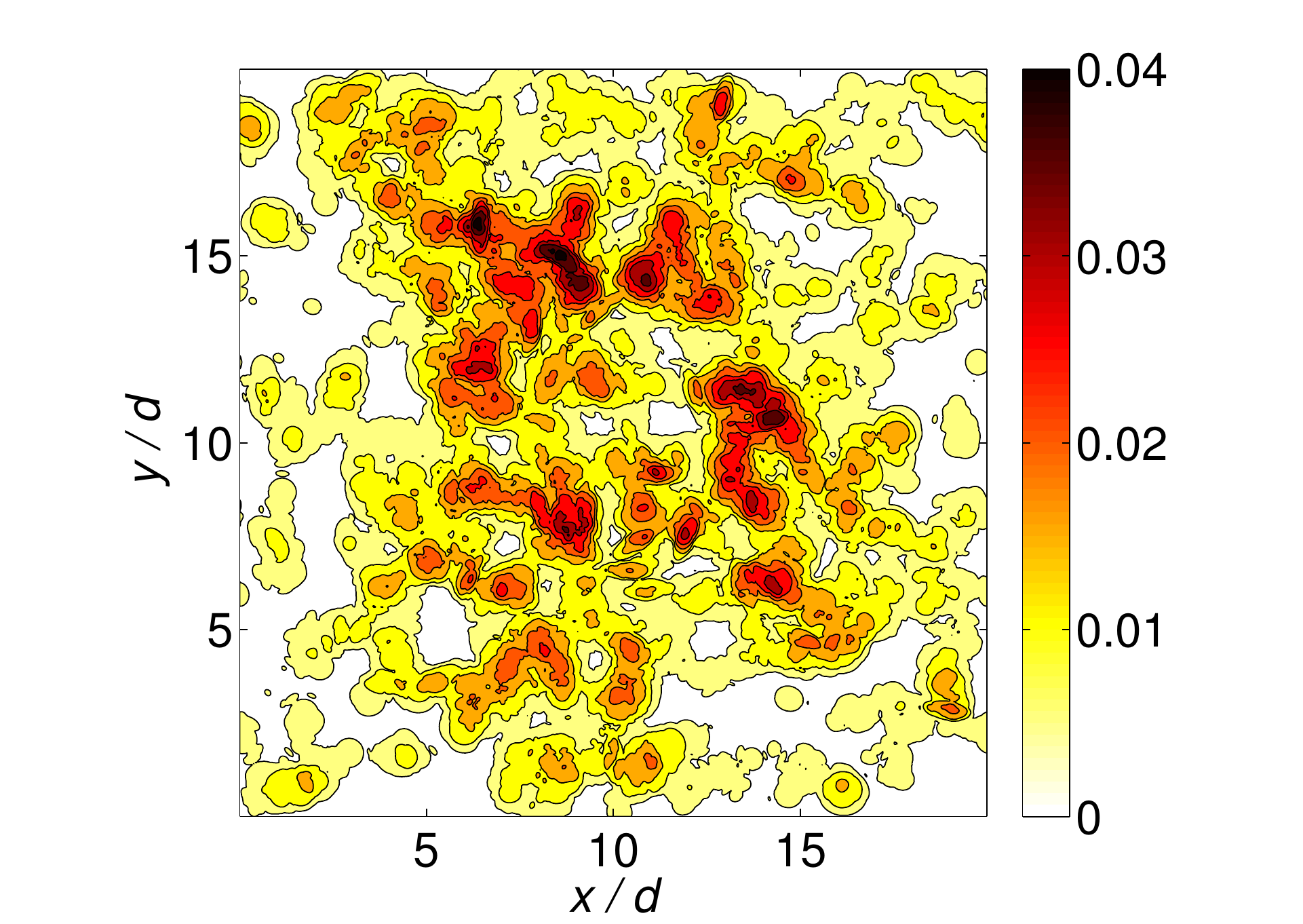}   
     \put(-188,120){{\large a)}}
     }%
  \subfigure{%
    \includegraphics[scale=0.34]{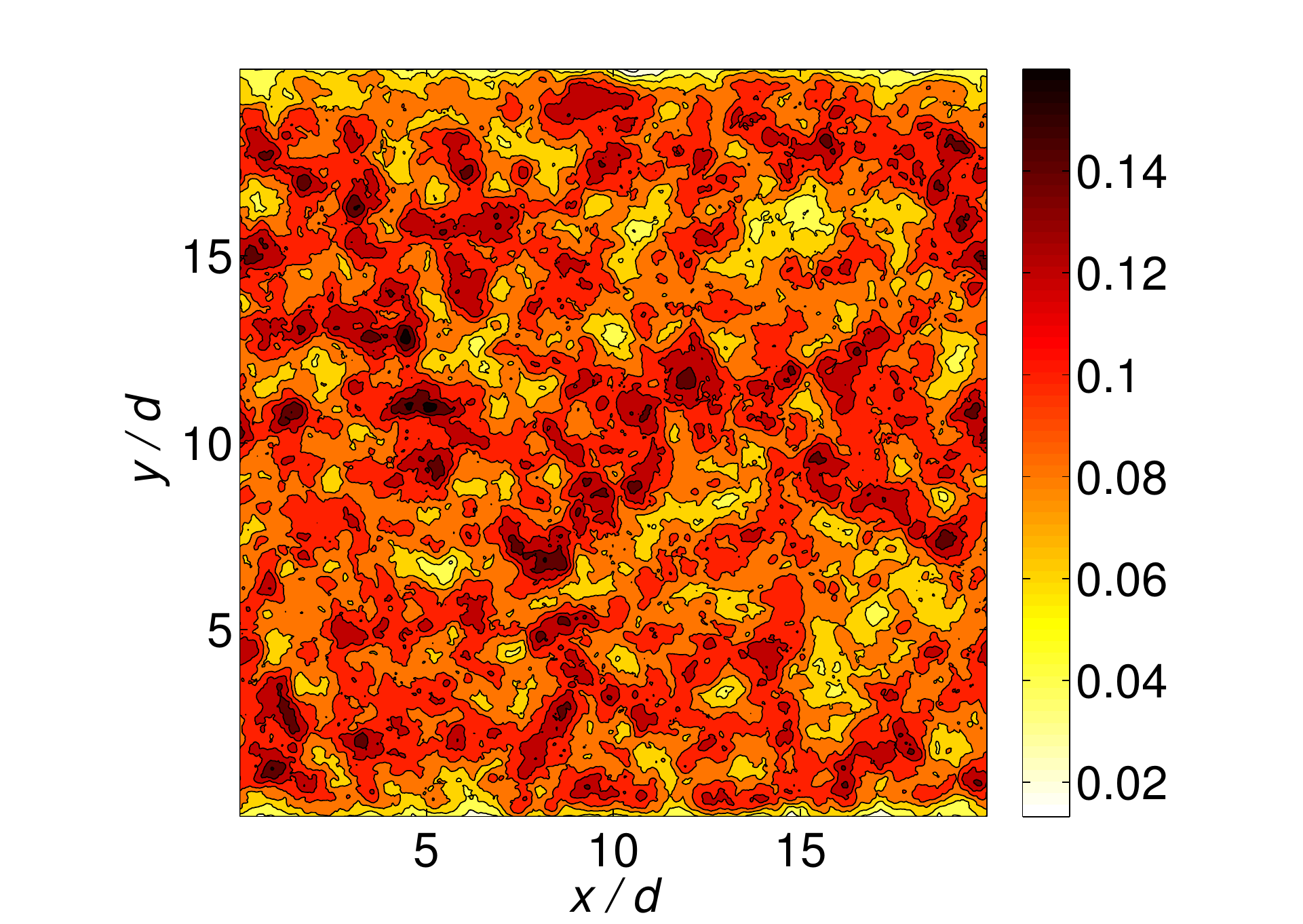}   
     \put(-188,120){{\large b)}}
     }\\%
  \subfigure{%
    \includegraphics[scale=0.34]{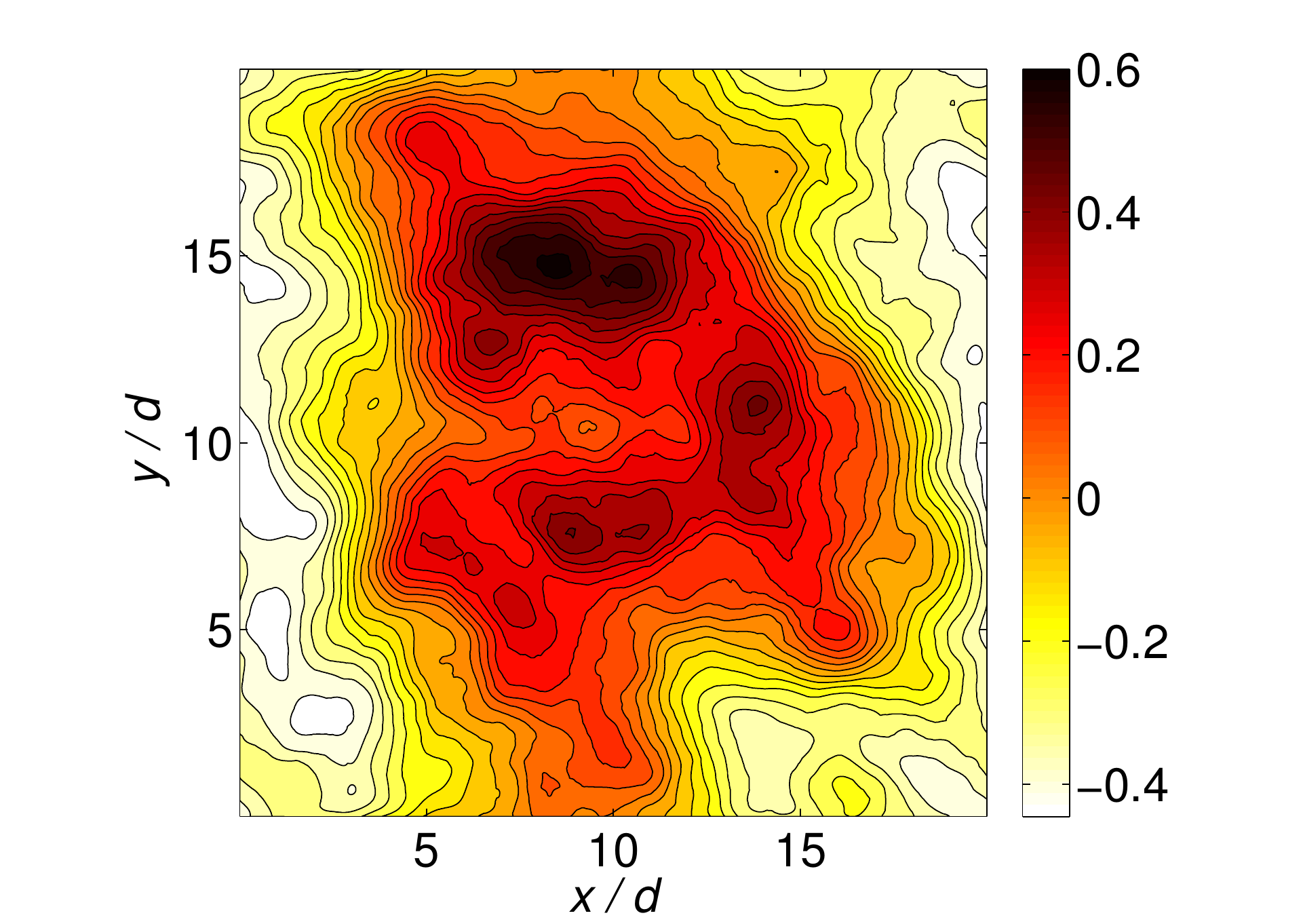}   
     \put(-188,120){{\large c)}}
     }%
  \subfigure{%
    \includegraphics[scale=0.34]{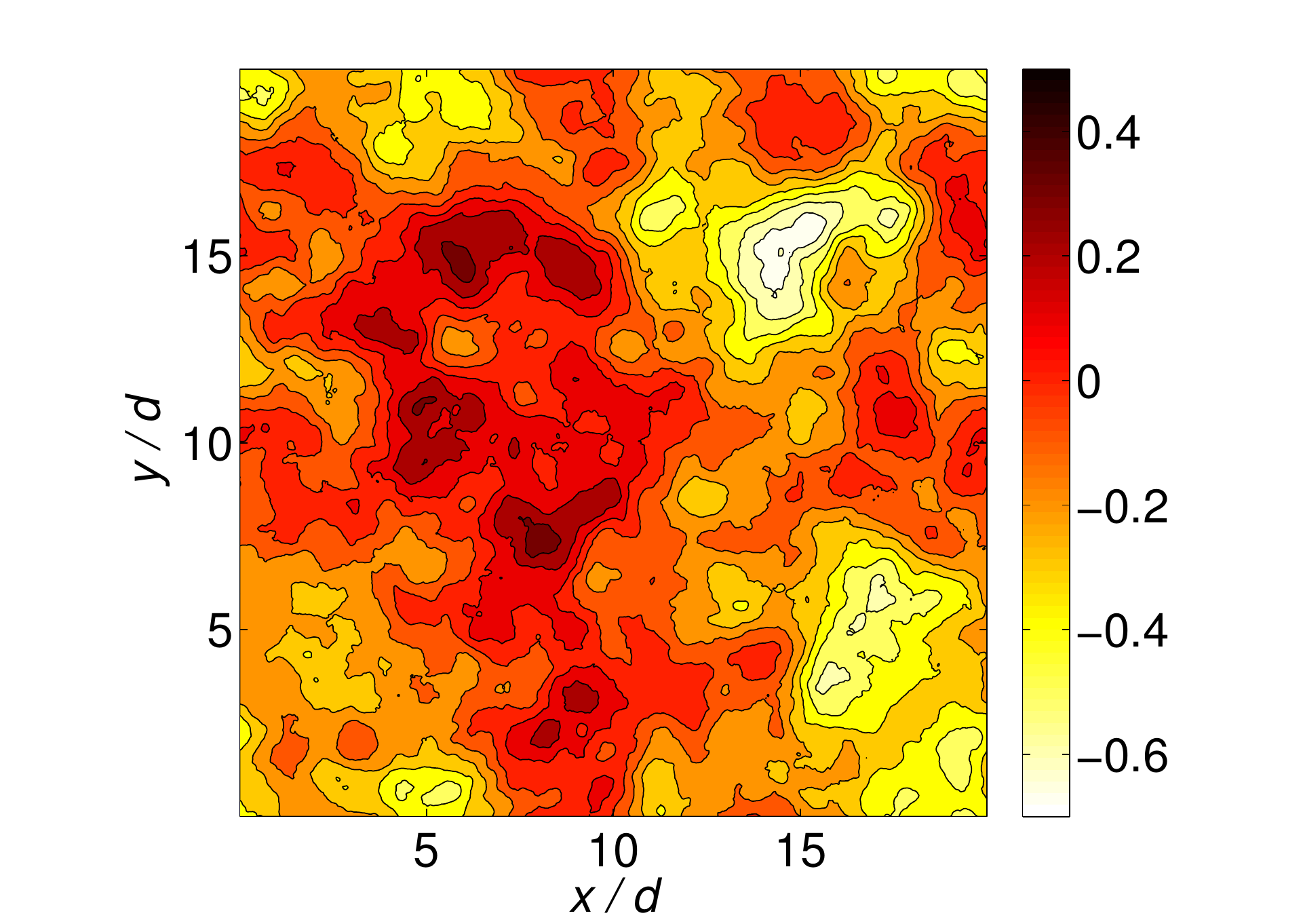}   
     \put(-188,120){{\large d)}}
     }%
\caption{Instantaneous particle concentration averaged over the settling ($z$-) direction for $\phi=1\%$ (a) and $10\%$ (b). 
Instantaneous vertical component of the fluid velocity averaged over the settling direction for $\phi=1\%$ (c) and $10\%$ (d).
These statistics are calculated at a single timestep towards the end of the simulation ($t \sim 182 \tau_p$ and $\sim 42 \tau_p$ 
for $\phi=1\%$ and $10\%$).}
\label{fig:flu_v}
\end{figure}

\subsection{Fluid phase velocity statistics}

Finally, we look at the statistics of the fluid-phase velocity. Figures~\ref{fig:flu_v}(a),(b) report the instantaneous 
particle concentration averaged over the settling direction, whereas figures~\ref{fig:flu_v}(c),(d) show the 
instantaneous vertical component of the fluid velocity, also averaged along the $z$- direction. For these statistics we choose a single 
timestep towards the end of the simulation ($t \sim 182 \tau_p$ and $\sim 42 \tau_p$ for $\phi=1\%$ and $10\%$). For $\phi=1\%$, we observe 
that most particles accumulate around the centre of the computational domain; see figure~\ref{fig:flu_v}(a). Within this columnar 
structure, particles settle substantially faster than $V_t$, and due to the no-slip boundary condition, the fluid surrounding the 
particles is forced to move in the same direction (positive $z$- direction). Indeed, in figure~\ref{fig:flu_v}(c) we see that the fluid 
speed $U_z$ is strongly positive in the same regions, with maxima at the locations of higher concentration. The highest fluid speed in the 
settling direction is almost of the order of the terminal velocity ($max (U_z) \simeq 0.6 V_t$). Note that due to the zero volume flux 
condition, in the locations depleted of particles the fluid moves in the direction opposite to gravity with non-negligible speeds ($min (U_z) 
\simeq -0.4$). This contributes to the hindrance effect. Similar results are found for $\phi=0.5\%$.

\begin{figure}
  \centering
  \subfigure{%
    \includegraphics[scale=0.34]{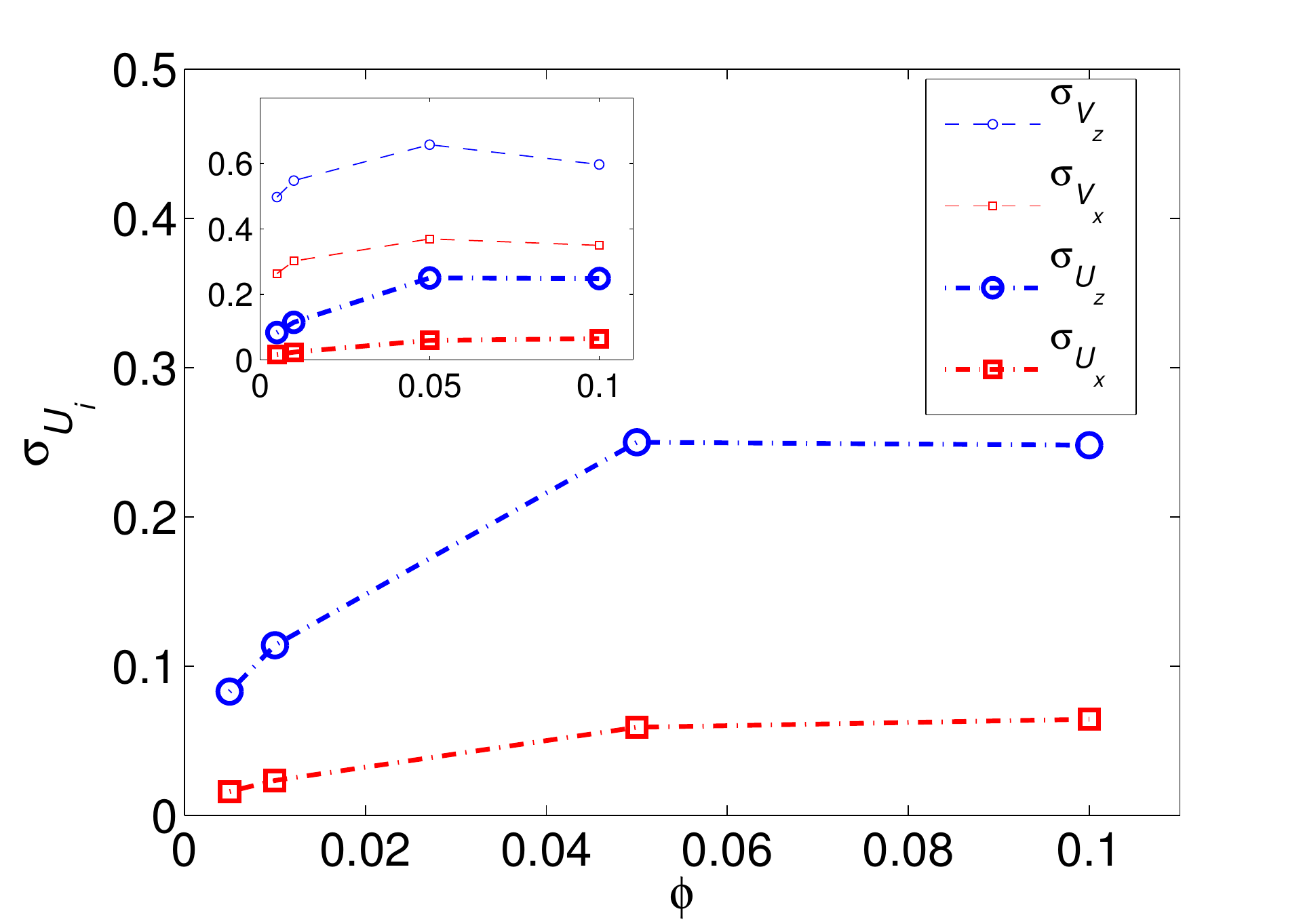}   
     \put(-188,120){{\large a)}}
     }%
  \subfigure{%
    \includegraphics[scale=0.34]{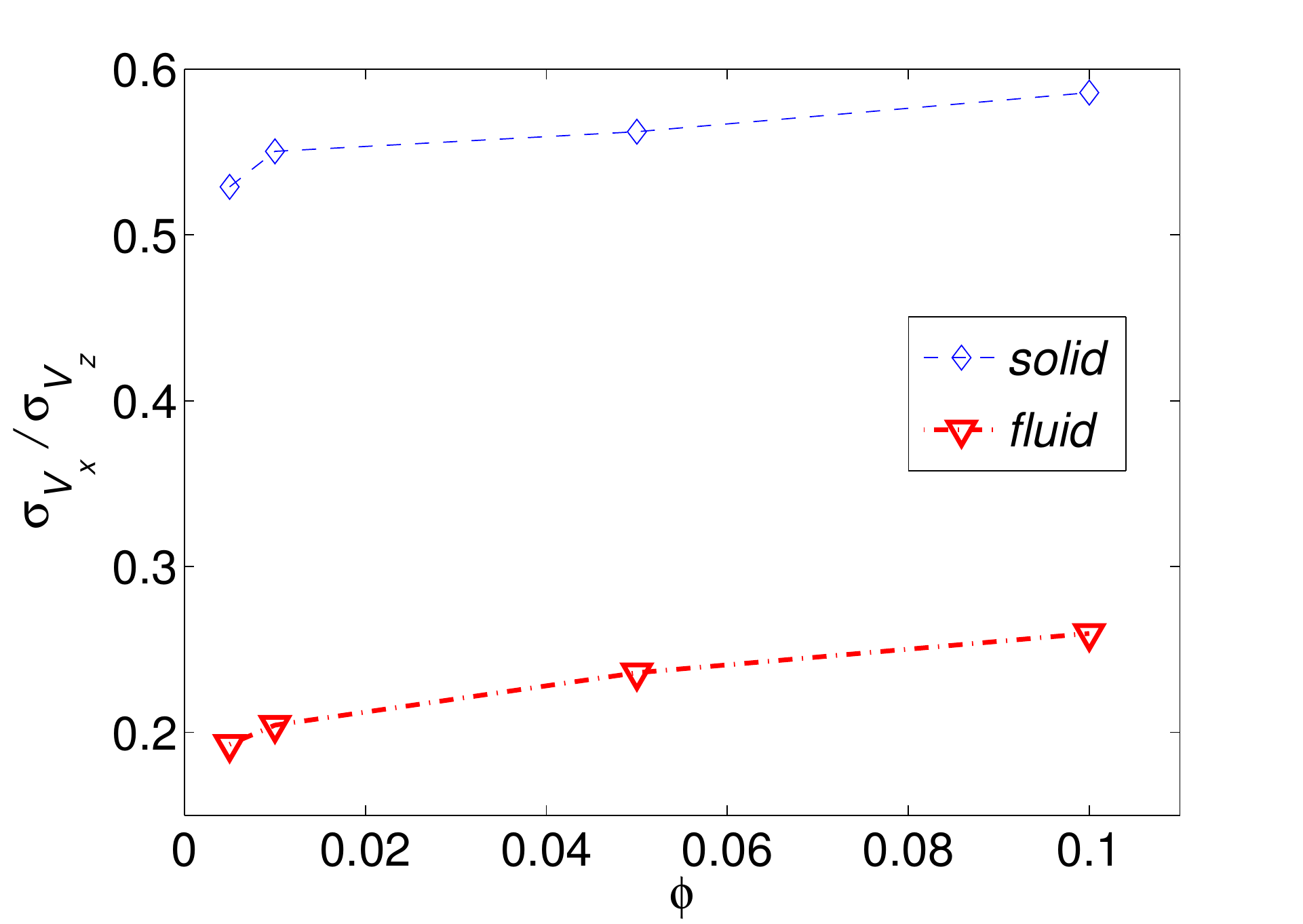}   
     \put(-188,120){{\large b)}}
     }%
\caption{(a) Standard deviation of the vertical and horizontal components of the fluid velocity, $\sigma_{U_z}$ and $\sigma_{U_x}$, for 
all $\phi$. In the inset, the standard deviation of the particles velocities are also shown. (b) Anisotropy of fluid velocity 
fluctuations, $\sigma_{U_x}/\sigma_{U_z}$, together with those of the solid phase, for all $\phi$.}
\label{fig:flu_v2}
\end{figure}

In figures~\ref{fig:flu_v}(b),(d) we show instead the mean particle concentration and mean fluid speed for $\phi=10\%$. The volume 
fraction is relatively high and particles are almost uniformly distributed in the domain. The locations of high positive fluid speed 
cannot be easily related to the positions of high concentration. However, we still observe that where the particle concentration is lower, 
the mean fluid speed is large and negative (i.e., rising fluid). So, for $\phi=10\%$ the maximum positive $U_z$ is reduced ($\simeq 0.3 
V_t$), while the rising fluid becomes faster ($min (U_z) \simeq -0.7$) leading to an increased hindrance effect.

Last, we show in figures~\ref{fig:flu_v2}(a),(b) the standard deviation of the vertical and horizontal components of the fluid 
velocity, $\sigma_{U_z}$ and $\sigma_{U_x}$, and the velocity fluctuations anisotropy, $\sigma_{U_x}/\sigma_{U_z}$, for all $\phi$. We 
see that both $\sigma_{U_z}$ and $\sigma_{U_x}$ increase with $\phi$. However, for $\sigma_{U_z}$ an approximately constant value of 
$0.25 V_t$ is reached after $\phi=5\%$. The standard deviations of the fluid velocities are substantially smaller than those of the particles, 
(especially in the horizontal direction), for the smaller volume fractions; see the inset of figure~\ref{fig:flu_v2}(a). The difference is 
progressively reduced as $\phi$ increases. For example, for $\phi=0.5\%$ we find that $\sigma_{U_z}=0.17 \sigma_{V_z}$, while for $\phi=10\%$, 
$\sigma_{U_z}=0.42 \sigma_{V_z}$. This indicates that at high volume fractions, the dynamics of both phases is governed mostly by excluded 
volume effects. Regarding the anisotropy of fluid velocity fluctuations, $\sigma_{U_x}/\sigma_{U_z}$, we see from figure~\ref{fig:flu_v2}(b) 
that it also increases with $\phi$. As for the solid phase, the increase with $\phi$ is almost linear.

\section{Final remarks}

We have studied the sedimentation of suspensions of oblate particles in quiescent fluid at finite $Re_t$. We chose 
the aspect ratio $\AR = 1/3$, density ratio $R=1.5$, Galileo number $Ga=60$ (based on the diameter of a sphere 
with equal volume), and four volume fractions $\phi=0.5\%, 1\%, 5\%$ and $10\%$. The single particle case was also 
simulated and at this combination of $R$ and $Ga$, the particle settles with its broad side perpendicular to the 
direction of gravity with a straight, steady wake. The orientation vector of the isolated oblate is hence $[O_x, 
O_y, O_z]=[0, 0, 1]$ (i.e. the pitch angle with respect to the plane perpendicular to gravity is $0^o$).

The average settling speed of a suspension changes with the particle volume fraction as the results of the competition of two 
different physical mechanisms: i) the hindrance effect, which is more pronounced in higher volume 
fractions and tends to reduce the average settling speed;  and
ii) the hydrodynamics of particle pair interactions, e.g. drafting-kissing-tumbling, which tends 
to increase the average speed and form piles of particles. We report that, unlike the case of 
spherical particles of equal $Ga$, the mean settling speed of the oblate particles suspension, 
$\langle V_z \rangle/V_t$, increases with $\phi$ in dilute conditions. For $\phi=0.5\%-1\%$; the mean settling speed is about 
$30\%$ larger than the terminal velocity of an isolated oblate. Note that for suspensions of spheres it has 
been shown that $\langle V_z \rangle/V_t$ is always a decreasing function of $\phi$ for $Ga \sim 60$ 
\citep{richardson1954,yin2007,uhlmann2014}. The mean settling speed becomes smaller than the terminal velocity 
$V_t$ only for $\phi > 5\%$. This implies that at lower volume fractions the hindrance effect is overcome by hydrodynamic and 
particle-particle interactions and this leads to an increase of $\langle V_z \rangle/V_t$. Indeed, we have shown that in dilute 
conditions most particles are arranged at steady-state in a columnar-like structure with a 
radius of about $20c$, where $c$ is the oblate polar radius. Within this structure, particle clusters settle with 
velocities up to $4$ times the mean, $\langle V_z \rangle/V_t$. Therefore, the probability density functions, $p.d.f.$s, of 
$V_z/V_t$ display a clear positive skewness, $S_{V_z} \sim 0.4$ (i.e. many particles fall with speeds larger than the mean 
value). 
While $\langle V_z \rangle/V_t$ is reduced increasing the volume fraction, the velocity standard deviation 
$\sigma_{V_z}$ increases up to $\phi \sim 5\%$, the skewness $S_{V_z}$ tends to $0$ and the flatness $F_{V_z}$ is 
always approximately $3$. Additionally, within the columnar-like structure, the fluid is strongly dragged by the 
particles (due to the no-slip condition) and in the settling direction it reaches speeds almost of the order of $V_t$ 
($max (U_z) \sim 0.6 V_t$).

To study the suspension microstructure, the pair- and radial distribution functions are calculated. While no 
clustering is observed for spheres of same $Ga$, the pair-distribution function $P(\vec r)$ is found to be large 
all around a reference particle and especially between $\psi \sim 0^o - 80^o$ and $r/c \sim 2 - 5$. The highest 
probability of finding a neighbour particle is at $r/c = 2.02$, $\psi \simeq 17^o$ for $\phi=0.5\%$, $r/c 
= 2.02$, $\psi \simeq 10^o$ for $\phi=1\%$, and around $r/c = 2.02$, $\psi \simeq 0^o-2^o$ for the highest volume fractions investigated. 
Hence, on average particles are almost piled up. The extent of clustering sharply decreases for $\phi > 
1\%$ as it is shown by the pair-distribution function $P(\vec r)$ and its angular average, $g(r)$ (the radial 
distribution function). The radial distribution function is found to be maximum around $r/c \simeq 4$ (i.e.\ regardless 
of the orientation, the highest probability of finding a particle-pair is located at a separation 
distance of $4c$). In the cases with lowest volume fractions, the particle distribution is found to be uncorrelated 
at distances larger than  $ 20c$, a distance of the order of the radius of the columnar structure. For the denser 
cases, the decorrelation of the structure occurs at shorter separations, at distances of the order of $6c$. Using the 
definition of order parameter $\langle P_2 \rangle (r)$ we have shown that for distances of about $2c$ particles are almost 
perfectly vertically aligned. Above $r/c > 3$ particles are almost horizontally aligned (with a finite inclination 
with respect to the horizontal plane). 
The suspension structure 
becomes more isotropic for distances $r/c \geq 6$ in the denser cases and $r/c \geq 8$ in the more dilute cases.

We have also considered the particle lateral velocity as well as the angular velocities. The mean 
particle lateral speed $\langle V_x \rangle/V_t$ is approximately $0$, while its standard deviation $\sigma_{V_x}$ 
increases with the volume fraction up to $\phi=5\%$ and is smaller than $\sigma_{V_z}$. 
The anisotropy of the particle velocity fluctuations 
$\sigma_{V_x}/\sigma_{V_z}$ increases abruptly until $\phi=0.5\%$ and then approximately linearly with 
$\phi$. From the $p.d.f.$s of angular velocities, we find that particles rotate more around the 
directions perpendicular to gravity. It is found that particles settle on average inclined with respect to 
the horizontal plane (the particle orientation with respect to the axis of symmetry, $|O_z|$, is less than $1$, where 1 
corresponds to a pitch angle of $0^o$). The mean pitch angle increases with $\phi$ from $22.8^o$ to $47^o$ 
(i.e.\ $|O_z|$ decreases with $\phi$). A power-law fit that depends solely on the solid volume fraction $\phi$ and one 
coefficient is proposed.
Computing  the terminal speed of an 
isolated oblate settling with the mean pitch of the suspension with $\phi=0.5\%$ 
 ($22.8^o$) we find that the lateral velocity $V_x/V_t=0.41$ and the vertical only $V_z/V_t=1.03$, 
 significantly lower than that of the suspension, showing the importance of particle-pair interactions. 

Finally, we have calculated the joint probability functions of particle settling speed and orientation. We used 
conditioned averages to show that particles settling with speeds larger than the mean $\langle V_z \rangle/V_t$, 
have on average lower mean orientations (higher pitch angles).

With this study we have began to look at the effects of particle shape in sedimenting suspensions of inertial particles. In the 
future, it will be interesting to consider oblate and prolate particles of different aspect ratios and Galileo numbers.

%
\begin{acknowledgments}
This work was supported by the European Research Council Grant No.\ ERC-2013-CoG-616186, TRITOS, and by the Swedish Research Council (VR).
Computer time provided by SNIC (Swedish
National Infrastructure for Computing) and the support from the COST Action MP1305: \emph{Flowing matter} are acknowledged.
\end{acknowledgments}

%


\begin{thebibliography}{38}
\expandafter\ifx\csname natexlab\endcsname\relax\def\natexlab#1{#1}\fi

\bibitem[Ardekani {\em et~al.\/}(2016)Ardekani, Costa, Breugem \&
  Brandt]{ardekani2016}
{\sc Ardekani, Mehdi~Niazi, Costa, Pedro, Breugem, Wim~Paul \& Brandt, Luca}
  2016 Numerical study of the sedimentation of spheroidal particles. {\em
  International Journal of Multiphase Flow\/} {\bf 87}, 16--34.

\bibitem[Batchelor(1972)]{batchelor1972}
{\sc Batchelor, GK} 1972 Sedimentation in a dilute dispersion of spheres. {\em
  Journal of Fluid Mechanics\/} {\bf 52}~(02), 245--268.

\bibitem[Bouchet {\em et~al.\/}(2006)Bouchet, Mebarek \&
  Du{\v{s}}ek]{bouchet2006}
{\sc Bouchet, G, Mebarek, M \& Du{\v{s}}ek, J} 2006 Hydrodynamic forces acting
  on a rigid fixed sphere in early transitional regimes. {\em European Journal
  of Mechanics-B/Fluids\/} {\bf 25}~(3), 321--336.

\bibitem[Breugem(2012)]{breugem2012}
{\sc Breugem, W-P} 2012 A second-order accurate immersed boundary method for
  fully resolved simulations of particle-laden flows. {\em J. Comput. Phys.\/}
  {\bf 231}~(13), 4469--4498.

\bibitem[Brosse \& Ern(2011)]{brosse2011}
{\sc Brosse, N \& Ern, P} 2011 Paths of stable configurations resulting from
  the interaction of two disks falling in tandem. {\em Journal of Fluids and
  Structures\/} {\bf 27}~(5), 817--823.

\bibitem[Chouippe \& Uhlmann(2015)]{chouippe2015}
{\sc Chouippe, Agathe \& Uhlmann, Markus} 2015 Forcing homogeneous turbulence
  in direct numerical simulation of particulate flow with interface resolution
  and gravity. {\em Physics of Fluids (1994-present)\/} {\bf 27}~(12), 123301.

\bibitem[Chrust(2012)]{chrust2012}
{\sc Chrust, Marcin} 2012 Etude num{\'e}rique de la chute libre d'objets
  axisym{\'e}triques dans un fluide newtonien. PhD thesis, Strasbourg.

\bibitem[Clift {\em et~al.\/}(2005)Clift, Grace \& Weber]{clift2005}
{\sc Clift, Roland, Grace, John~R \& Weber, Martin~E} 2005 {\em Bubbles, drops,
  and particles\/}. Courier Corporation.

\bibitem[Climent \& Maxey(2003)]{climent2003}
{\sc Climent, E \& Maxey, MR} 2003 Numerical simulations of random suspensions
  at finite Reynolds numbers. {\em International Journal of Multiphase Flow\/}
  {\bf 29}~(4), 579--601.

\bibitem[Costa {\em et~al.\/}(2015)Costa, Boersma, Westerweel \&
  Breugem]{costa2015}
{\sc Costa, Pedro, Boersma, Bendiks~Jan, Westerweel, Jerry \& Breugem,
  Wim-Paul} 2015 Collision model for fully resolved simulations of flows laden
  with finite-size particles. {\em Physical Review E\/} {\bf 92}~(5), 053012.

\bibitem[Di~Felice(1999)]{di1999}
{\sc Di~Felice, R} 1999 The sedimentation velocity of dilute suspensions of
  nearly monosized spheres. {\em International Journal of Multiphase Flow\/}
  {\bf 25}~(4), 559--574.

\bibitem[Doostmohammadi \& Ardekani(2015)]{doost2015}
{\sc Doostmohammadi, A \& Ardekani, AM} 2015 Suspension of solid particles in a
  density stratified fluid. {\em Physics of Fluids (1994-present)\/} {\bf
  27}~(2), 023302.

\bibitem[Ern {\em et~al.\/}(2012)Ern, Risso, Fabre \& Magnaudet]{ern2012}
{\sc Ern, Patricia, Risso, Fr{\'e}d{\'e}ric, Fabre, David \& Magnaudet,
  Jacques} 2012 Wake-induced oscillatory paths of bodies freely rising or
  falling in fluids. {\em Annual Review of Fluid Mechanics\/} {\bf 44},
  97--121.

\bibitem[Feng {\em et~al.\/}(1994)Feng, Hu \& Joseph]{feng1994}
{\sc Feng, James, Hu, Howard~H \& Joseph, Daniel~D} 1994 Direct simulation of
  initial value problems for the motion of solid bodies in a newtonian fluid
  part 1. sedimentation. {\em Journal of Fluid Mechanics\/} {\bf 261}, 95--134.

\bibitem[Fonseca \& Herrmann(2005)]{fonseca}
{\sc Fonseca, F \& Herrmann, HJ} 2005 Simulation of the sedimentation of a
  falling oblate ellipsoid. {\em Physica A: Statistical Mechanics and its
  Applications\/} {\bf 345}~(3), 341--355.

\bibitem[Fornari {\em et~al.\/}(2016{\natexlab{{\em a\/}}})Fornari, Picano \&
  Brandt]{fornari2015}
{\sc Fornari, W, Picano, F \& Brandt, L} 2016{\natexlab{{\em a\/}}}
  Sedimentation of finite-size spheres in quiescent and turbulent environments.
  {\em J. Fluid Mech.\/} {\bf 788}, 640--669.

\bibitem[Fornari {\em et~al.\/}(2016{\natexlab{{\em b\/}}})Fornari, Picano,
  Sardina \& Brandt]{fornari2016b}
{\sc Fornari, Walter, Picano, Francesco, Sardina, Gaetano \& Brandt, Luca}
  2016{\natexlab{{\em b\/}}} Reduced particle settling speed in turbulence.
  {\em Journal of Fluid Mechanics\/} {\bf 808}, 153--167.

\bibitem[Fortes {\em et~al.\/}(1987)Fortes, Joseph \& Lundgren]{fortes1987}
{\sc Fortes, Antonio~F, Joseph, Daniel~D \& Lundgren, Thomas~S} 1987 Nonlinear
  mechanics of fluidization of beds of spherical particles. {\em J. Fluid
  Mech.\/} {\bf 177}, 467--483.

\bibitem[Garside \& Al-Dibouni(1977)]{garside1977}
{\sc Garside, John \& Al-Dibouni, Maan~R} 1977 Velocity-voidage relationships
  for fluidization and sedimentation in solid-liquid systems. {\em Industrial
  \& engineering chemistry process design and development\/} {\bf 16}~(2),
  206--214.

\bibitem[Guazzelli \& Morris(2011)]{guazzelli2011}
{\sc Guazzelli, Elisabeth \& Morris, Jeffrey~F} 2011 {\em A physical
  introduction to suspension dynamics\/}. Cambridge University Press.

\bibitem[Hasimoto(1959)]{hasimoto1959}
{\sc Hasimoto, H} 1959 On the periodic fundamental solutions of the stokes
  equations and their application to viscous flow past a cubic array of
  spheres. {\em Journal of Fluid Mechanics\/} {\bf 5}~(02), 317--328.

\bibitem[Horowitz \& Williamson(2010)]{horowitz2010}
{\sc Horowitz, M \& Williamson, CHK} 2010 The effect of Reynolds number on the
  dynamics and wakes of freely rising and falling spheres. {\em Journal of
  Fluid Mechanics\/} {\bf 651}, 251--294.

\bibitem[Huisman {\em et~al.\/}(2016)Huisman, Barois, Bourgoin, Chouippe,
  Doychev, Huck, Morales, Uhlmann \& Volk]{huisman2016}
{\sc Huisman, Sander~G, Barois, Thomas, Bourgoin, Micka{\"e}l, Chouippe,
  Agathe, Doychev, Todor, Huck, Peter, Morales, Carla E~Bello, Uhlmann, Markus
  \& Volk, Romain} 2016 Columnar structure formation of a dilute suspension of
  settling spherical particles in a quiescent fluid. {\em Physical Review
  Fluids\/} {\bf 1}~(7), 074204.

\bibitem[Jeffrey(1982)]{jeffrey1982}
{\sc Jeffrey, DJ} 1982 Low-Reynolds-number flow between converging spheres.
  {\em Mathematika\/} {\bf 29}~(1), 58--66.

\bibitem[Jenny {\em et~al.\/}(2004)Jenny, Du{\v{s}}ek \& Bouchet]{jenny2004}
{\sc Jenny, Mathieu, Du{\v{s}}ek, J \& Bouchet, G} 2004 Instabilities and
  transition of a sphere falling or ascending freely in a newtonian fluid. {\em
  Journal of Fluid Mechanics\/} {\bf 508}, 201--239.

\bibitem[Kulkarni \& Morris(2008)]{kulkarni2008}
{\sc Kulkarni, Pandurang~M \& Morris, Jeffrey~F} 2008 Suspension properties at
  finite Reynolds number from simulated shear flow. {\em Physics of Fluids\/}
  {\bf 20}~(4), 040602.

\bibitem[Lambert {\em et~al.\/}(2013)Lambert, Picano, Breugem \&
  Brandt]{lambert2013}
{\sc Lambert, R~A, Picano, F, Breugem, W-P \& Brandt, L} 2013 Active
  suspensions in thin films: nutrient uptake and swimmer motion. {\em J. Fluid
  Mech.\/} {\bf 733}, 528--557.

\bibitem[Magnaudet \& Mougin(2007)]{magnaudet2007}
{\sc Magnaudet, Jacques \& Mougin, Guillaume} 2007 Wake instability of a fixed
  spheroidal bubble. {\em Journal of Fluid Mechanics\/} {\bf 572}, 311--337.

\bibitem[Picano {\em et~al.\/}(2015)Picano, Breugem \& Brandt]{picano2015}
{\sc Picano, F, Breugem, W-P \& Brandt, L} 2015 Turbulent channel flow of dense
  suspensions of neutrally buoyant spheres. {\em J. Fluid Mech.\/} {\bf 764},
  463--487.

\bibitem[Richardson \& Zaki(1954)]{richardson1954}
{\sc Richardson, JF \& Zaki, WN} 1954 The sedimentation of a suspension of
  uniform spheres under conditions of viscous flow. {\em Chemical Engineering
  Science\/} {\bf 3}~(2), 65--73.

\bibitem[Sangani \& Acrivos(1982)]{sangani1982}
{\sc Sangani, AS \& Acrivos, A} 1982 Slow flow past periodic arrays of
  cylinders with application to heat transfer. {\em International journal of
  Multiphase flow\/} {\bf 8}~(3), 193--206.

\bibitem[Santarelli \& Fr{\"o}hlich(2015)]{santarelli2015}
{\sc Santarelli, C \& Fr{\"o}hlich, J} 2015 Direct numerical simulations of
  spherical bubbles in vertical turbulent channel flow. {\em International
  Journal of Multiphase Flow\/} {\bf 75}, 174--193.

\bibitem[Santarelli \& Fr{\"o}hlich(2016)]{santarelli2016}
{\sc Santarelli, C \& Fr{\"o}hlich, J} 2016 Direct numerical simulations of
  spherical bubbles in vertical turbulent channel flow. influence of bubble
  size and bidispersity. {\em International Journal of Multiphase Flow\/} {\bf
  81}, 27--45.

\bibitem[Schiller \& Naumann(1935)]{schil1935}
{\sc Schiller, L \& Naumann, A} 1935 A drag coefficient correlation. {\em Vdi
  Zeitung\/} {\bf 77}~(318), 51.

\bibitem[Tanaka \& Teramoto(2015)]{tanaka2015}
{\sc Tanaka, M \& Teramoto, D} 2015 Modulation of homogeneous shear turbulence
  laden with finite-size particles. {\em Journal of Turbulence\/} {\bf
  16}~(10), 979--1010.

\bibitem[Uhlmann \& Doychev(2014)]{uhlmann2014}
{\sc Uhlmann, M \& Doychev, T} 2014 Sedimentation of a dilute suspension of
  rigid spheres at intermediate galileo numbers: the effect of clustering upon
  the particle motion. {\em J. Fluid Mech.\/} {\bf 752}, 310--348.

\bibitem[Yin \& Koch(2007)]{yin2007}
{\sc Yin, X \& Koch, D~L} 2007 Hindered settling velocity and microstructure in
  suspensions of solid spheres with moderate Reynolds numbers. {\em Phys
  Fluids\/} {\bf 19}~(9), 093302.

\bibitem[Zaidi {\em et~al.\/}(2014)Zaidi, Tsuji \& Tanaka]{zaidi2014}
{\sc Zaidi, Ali~Abbas, Tsuji, Takuya \& Tanaka, Toshitsugu} 2014 Direct
  numerical simulation of finite sized particles settling for high Reynolds
  number and dilute suspension. {\em International Journal of Heat and Fluid
  Flow\/} {\bf 50}, 330--341.

\end{thebibliography}

\end{document}